# Sulfur Vacancy Related Optical Transitions in Graded Alloys of $Mo_xW_{1-x}S_2$ Monolayers

Mahdi Ghafariasl [1], Tianyi Zhang [2], Zachary D. Ward [3], Da Zhou [4], David Sanchez [2], Venkataraman Swaminathan [5], Humberto Terrones [3], Mauricio Terrones [2,4,6], Yohannes Abate [1*]

[1] Department of Physics and Astronomy, University of Georgia, Athens, Georgia 30602, USA

[2] Department of Materials Science and Engineering, The Pennsylvania State University, University Park, PA 16802, USA

[3] Department of Physics and Astronomy, Rensselaer Polytechnic Institute, Rensselaer, New York 12180, USA

[4] Department of Physics, The Pennsylvania State University, University Park, PA 16802, USA

[5] Department of Materials Science and Nanoengineering, Rice University, Houston, TX 77005, USA

[6] Department of Chemistry, The Pennsylvania State University, University Park, PA 16802, USA

*Corresponding author E-mail: yohannes.abate@uga.edu

## ABSTRACT

Engineering the electronic bandgap is of utmost importance in diverse domains ranging from information processing and communication technology to sensing and renewable energy applications. Transition metal dichalcogenides (TMDCs) provide an ideal platform for achieving this goal through techniques including alloying, doping, and creating in-plane or out-of-plane heterostructures. Here, we report on the synthesis and characterization of atomically controlled two-dimensional graded alloy of $Mo_xW_{1-x}S_2$, wherein the center region is Mo rich and gradually transitions towards a higher concentration of W atoms at the edges. This unique alloy structure leads to a continuously tunable bandgap, ranging from 1.85 eV in the center to 1.95 eV at the edges consistent with the larger band gap of $WS_2$ relative to $MoS_2$. Aberration-corrected high-angle annular dark-field scanning transmission electron microscopy showed the presence of sulfur monovacancy, $V_S$, whose concentration varied across the graded $Mo_xW_{1-x}S_2$ layer as a function of Mo content with the highest value in the Mo rich center region. Optical spectroscopy measurements supported by *ab initio* calculations reveal a doublet electronic state of $V_S$, which was split due to the spin-orbit interaction, with energy levels close to the conduction band or deep in the band gap depending on whether the vacancy is surrounded by W atoms or Mo atoms. This unique electronic configuration of $V_S$ in the alloy gave rise to four spin-allowed optical transitions between the $V_S$ levels and the valence bands. Our work highlights the potential of simultaneous defect and optical engineering of novel devices based on these 2D monolayers.

## INTRODUCTION

Bandgap engineering is of great importance in modern semiconductor technology because of its capability for tuning the materials' optical and electrical properties that are key to their applications. In the past decade, the emergence of atomically-thin two-dimensional (2D) layered materials have provided an open canvas to engineer their bandgaps for device applications such as tunable lasers, light-emitting diodes (LEDs), and nanoelectronics[1,2]. Among the family of 2D materials [3-5] , monolayers of semiconducting transition metal dichalcogenides (TMDCs), such as $MoS_2$ and $WSe_2$ [6], possess direct bandgaps at optical frequencies, making them excellent candidates for bandgap engineering via diverse approaches such as alloying, hetero-stacking, strain engineering, intercalation, temperature control, and applying external electric fields[7-15] . Compared to bandgap tuning approaches that rely on external applied factors (e.g., temperature, strain, and electric field), alloying provides an effective and stable control because the bandgap is controlled by the intrinsic chemical composition, and a continuous bandgap engineering can be achieved by precisely modulating the alloy composition [16,17]. To date, ternary alloys of monolayer TMDCs (e.g., $Mo_xW_{1-x}S_2$, $MoS_{2x}Se_{2(1-x)}$) have been achieved by means of chemical vapor deposition (CVD)[18,19], physical vapor deposition (PVD)[20,21], exfoliation of bulk crystals synthesized by chemical vapor transport (CVT) [22], and other methods, allowing for the access of tunable optical properties as a function of compositions.

We employed an alkali metal halide-assisted CVD method to synthesize alloyed $Mo_XW_{1-X}S_2$ monolayers that showcase intriguing compositional gradients within individual crystals, and we thoroughly examined their optical properties. The alloys exhibit distinct compositional gradient transitioning from a Mo-rich center to a W-rich periphery. This strategically controlled alloying within a single crystal enabled the spatial tuning of the bandgap in a wide range (by over 0.1 eV), characterized by a uniform radial emission profile spanning across the entire flake, and facilitated the investigation of continuously composition-dependent intralayer optical transitions. Examining the defect structure in the alloy using aberration-corrected high-angle annular dark-field scanning transmission electron microscopy (AC-HAADF-STEM) showed the presence of sulfur monovacancy, $V_S$, whose concentration varied across the graded $Mo_xW_{1-x}S_2$ layer as a function of Mo content with the highest value in the Mo rich center region and the lowest value in the W rich edges. Detailed spectral analysis of the photoluminescence from the alloy as a function of temperature (4K-300K) and laser excitation intensity suggested several intralayer optical transitions besides the intrinsic band edge excitons. To investigate the origin of these transitions, detailed *ab initio* calculations were performed of the band structure of the alloy considering spin-orbit interactions. A doublet electronic state of $V_S$, which was split due to the spin-orbit interaction, was identified for the first time with energy levels close to the conduction band or deep in the band gap depending on whether the vacancy in the alloy is surrounded by W atoms or Mo atoms. This unique electronic configuration of $V_S$ in the alloy gave rise to four spin-allowed optical transitions between the $V_S$ levels and the valance bands. Matching the calculated transition energies with the peak positions of the deconvoluted peaks, enabled the identification of free-to-bound transitions involving a photoexcited electron captured at the doublet $V_S$ level and a hole in the top of the valence band at k=0. In addition, bound exciton transitions associated with the $V_S$ doublet were also identified. Thus, a multitude of $V_S$ related intralayer optical transitions reported for the first time in the alloy $Mo_xW_{1-x}S_2$, reveals the complex interplay between composition and defect structure thereby providing the exciting opportunity of combined defect and optical engineering to realize novel devices in alloyed TMDCs.

## RESULTS AND DISCUSSION

### Sample Synthesis and characterization

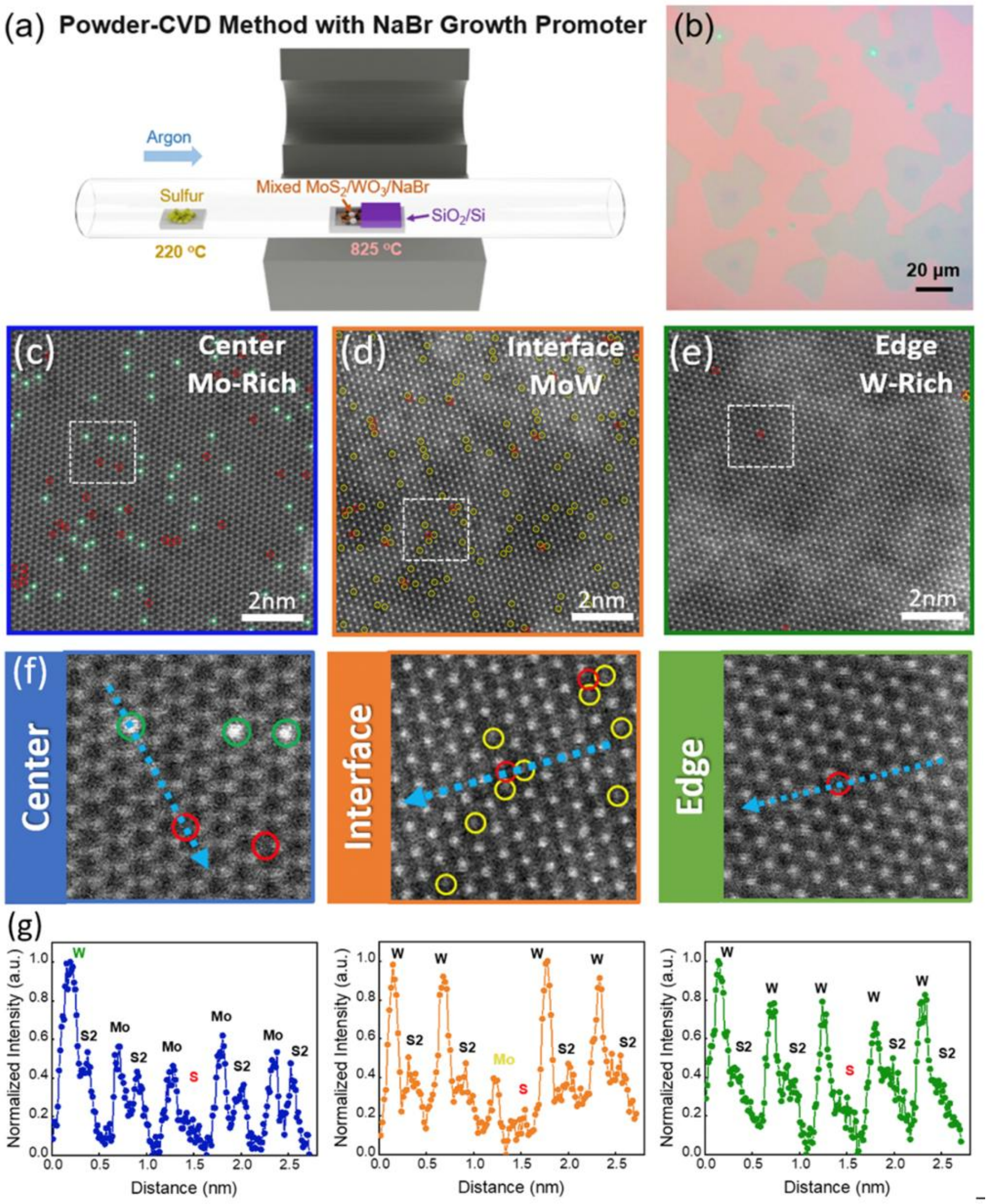

**Figure1.** The synthesis of Alloyed $Mo_xW_{1-x}S_2$ and AC-HRSTEM Structural Characterization (a) A schematic of the synthesis of alloyed $Mo_xW_{1-x}S_2$ monolayers with center-to-edge composition gradient. (b) Optical image of as-grown $Mo_XW_{1-X}S_2$ monolayer on $SiO_2$/Si substrate. (c−e) Typical AC-HAADF-STEM images acquired from the center, interface, and edge areas of the $Mo_XW_{1-X}S_2$ monolayer. The red, green and yellow circles denote the sulfur monovacancy ($V_S$), W atoms, and Mo atoms, respectively. (f) magnified images of the white dash boxes at the center, interface, and edge areas of the $Mo_XW_{1-X}S_2$ monolayer. (g) The corresponding intensity line profiles of areas marked by the blue arrows in (f) for three different regions of the alloyed $Mo_XW_{1-X}S_2$ monolayer.

Additive-assisted CVD has recently emerged as a powerful method for the preparation of 2D TMDCs and their heterostructures [23] .The use of growth additives, such as alkali metal halides (e.g., NaCl, NaBr) and organic compounds with aromatic structures (e.g., reduced graphene oxide (rGO), perylene-3,4,9,10-tetracarboxylic acid tetra potassium salt (PTAS), can effectively promote the growth of TMDCs with enhanced yield, increased grain sizes, and improved uniformity of layer numbers and morphology [24-26]. In our work, we employed an additive-assisted synthesis technique to prepare single-crystalline monolayers of alloyed $Mo_xW_{1-x}S_2$. An alkali metal halide-assisted CVD method was applied, with mixed $MoS_2$ and $WO_3$ powders as transition metal precursors and NaBr as a growth promoter (Figure 1(a)). An optical image of typical as-grown $Mo_xW_{1-x}S_2$ alloys is displayed in Fig. 1(b), showing truncated-triangular morphologies with noticeable optical contrast differences from center to edge regions. Unlike the in-plane heterostructures synthesized by liquid-phase precursor-assisted approach, which we previously reported [27,28], this method produced alloyed $Mo_xW_{1-x}S_2$ monolayers with center-to-edge composition gradient (a schematic is shown in Fig. 2(a)), similar to graded TMDC alloys synthesized previously[29-31]. Z-contrast aberration-corrected high angle annular dark field scanning transmission electron microscopy (AC-HAADF-STEM) imaging of the $Mo_xW_{1-x}S_2$ monolayer revealed that the Mo and W concentrations varied continuously from the center to the edge of the flake. The center regions contain a higher Mo concentration compared to the edge regions which are dominated by W with a compositional gradient interface between the two regions. This observation was further confirmed using far-field Raman and photoluminescence (PL) studies as discussed below. AC-HAADF-STEM imaging further identified that $V_S$ is the prevalent point defect in alloyed $Mo_xW_{1-x}S_2$ monolayers, which is labeled by the red-colored circles in Figs. 1(c)-(e). The center, interface, and edge regions contain estimated $V_S$ defect densities of 0.185 $nm^{-2}$, 0.091 $nm^{-2}$, 0.023 $nm^{-2}$, respectively. We note that there is a relationship between the prevalent transition metal (Mo or W) and the density of $V_S$. The center, which has the highest Mo concentration, also has the highest density of $V_S$. In contrast, the edge, which has the highest W concentration, has the lowest density of $V_S$. In the interface region, between the center and edge regions, we observed a $V_S$ density bounded by the two regions. The $V_S$ densities for the Mo-rich center and W-rich edge regions are in accordance with defect densities previously measured in CVD-grown $MoS_2$ and $WS_2$ monolayers [32,33].

**Optical Characterization of alloyed $Mo_XW_{1-X}S_2$ at room temperature**

To characterize the structural and optical properties of as-synthesized TMDC alloys, we performed far-field Raman and PL measurements. Figures 2(b) and 2(e) illustrate the Raman spectra acquired from different regions of $Mo_xW_{1-x}S_2$ alloys. The material exhibits Raman signatures that are consistent with the previous results, with $MoS_2$-like vibrational modes dominating in center regions and $WS_2$-like vibrational modes in edge regions, while the interface/middle regions exhibit a combination of both sets of modes [30,27]. The PL emissions from the center (~1.85 eV) and edge regions (~1.95 eV) of the alloyed structure are close to the optical band gap of pristine $MoS_2$ and $WS_2$, respectively (Fig. 2(c)).

To further elucidate the structural and optical property differences of the sample as a function of position within the flake, Raman and PL line scans were performed on a representative alloyed monolayer $Mo_xW_{1-x}S_2$ flake along the scan direction marked in Fig. 2(d) and plotted in Figs. 2(e) and (f). The intensities of both the convoluted $WS_2$ E' and 2LA(M) modes (black line) and the $MoS_2$ E' mode (red line) display a gradual change along the scan direction (Fig. 2(e)), which unambiguously indicates a composition gradient from center to edge of the flake. Consequently, the optical band gap of the alloyed $Mo_xW_{1-x}S_2$ was continuously modulated due to the lateral variation of the degree of alloying, indicated by the gradual change in PL peak positions (Fig. 2(f)) that is consistent with previous results on graded TMDC alloys [29,34]. We also map the band edge emission spatial profile of the $Mo_xW_{1-x}S_2$ alloy by taking hyperspectral PL imaging [28]. Figure 2(g) shows 3D hyperspectral data cube taken by measuring an array of 75 by 75 pixels PL normalized spectra of the alloyed monolayer. The x and y axes of the 3D data cube shown in Fig. 2(g) indicate the plane of the sample surface while the z-axis corresponds to the photon energy axis (1.85 to 2.0 eV). The acquisition time for each spectrum was 1 sec, and total acquisition time of 2 h per image (see Materials and Methods for details). The PL spatial map at a fixed energy is extracted by cross-section cut of the cube as shown in Fig. 2(h).

The energy dependent PL emission maps reveal 2D quasi-symmetric spatial variation of the degree of alloying from the center to the edge of the flake. In general, the energy gap of an alloy $A_xB_{1-x}$ in terms of the pure compound energy gap $E_A$ and $E_B$, follows an equation[35]

$$E(x) = E_B + (E_A - E_B - b)x + bx^2 \qquad (1)$$

Where $b$ is the bowing parameter. For $Mo_XW_{1-X}S_2$ fitting PL peak positions to the above equation Chen et al. [30] , obtained a b value of 0.25 eV for the A exciton peak and 0.19 eV for the B exciton peak. However, for the alloys $AB_{2(1-x)}C_{2x}$ where both B and C are chalcogen atoms, the bowing was found to be considerably smaller [17]. As the alloy composition is not directly determined in our samples, treating the scan position as a variable, and fitting the PL peaks in Fig. 2(f) to the above equation gives a value of b ~ 0.054 eV suggesting small bowing. From this we surmise that the small bowing parameter indicates small lattice mismatch/strain and thermodynamic miscibility in our samples due to the unique CVD sample synthesis method employed in our study.

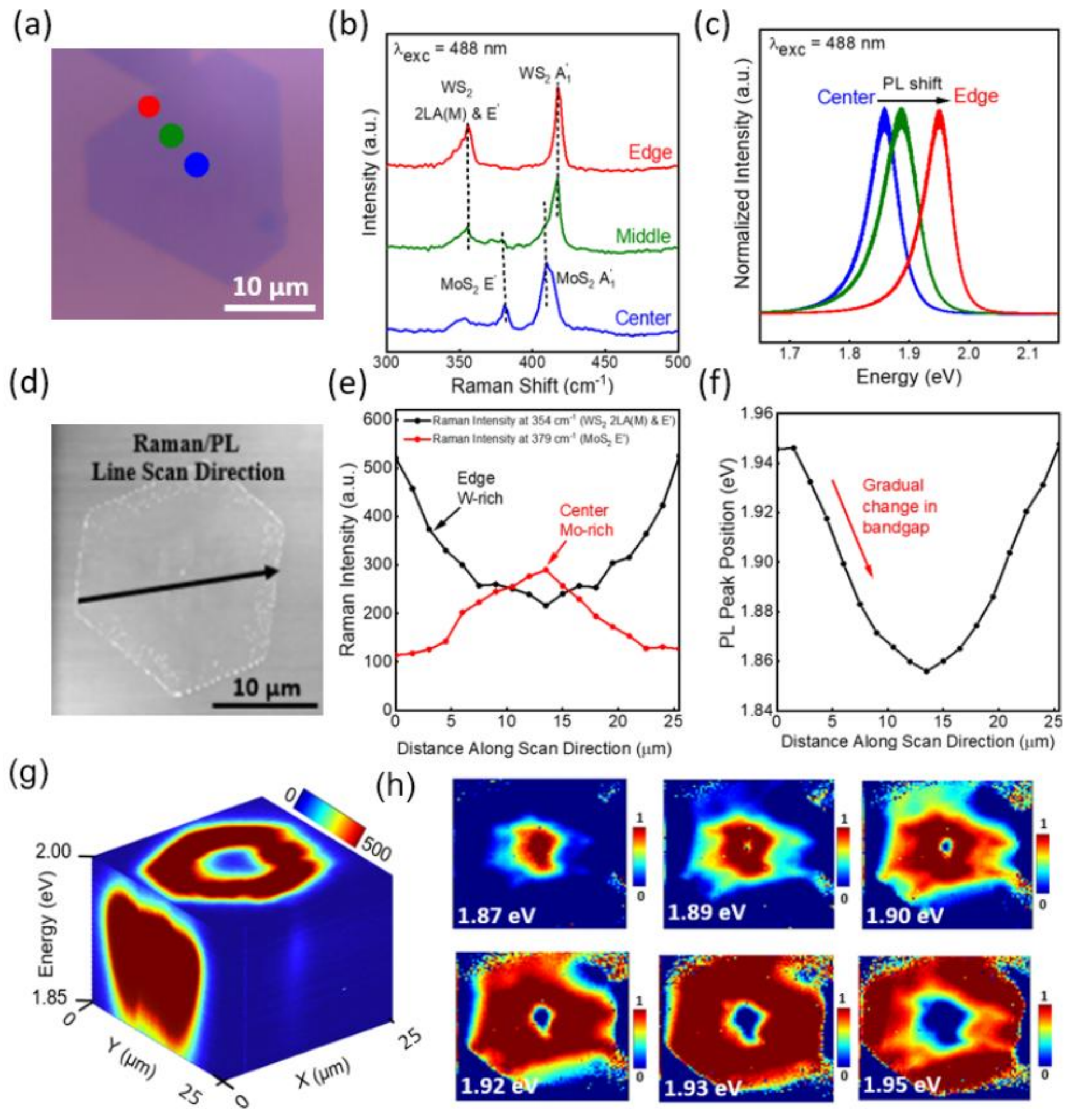

**Figure 2.** Optical Characterization of alloyed monolayer $Mo_xW_{1-x}S_2$. (a) Schematic of the structure of alloyed monolayer $Mo_xW_{1-x}S_2$. (b) Raman and (c) PL spectra of various regions in alloyed monolayer $Mo_xW_{1-x}S_2$. (d) Optical microscopy image of the $Mo_xW_{1-x}S_2$ flake used for Raman and PL line scans (the scan direction is marked by the black arrow). (e) Raman intensity profiles of $WS_2$ E' and 2LA(M) modes (black) and $MoS_2$ E' mode (red line) as a function of distance along the scan direction. Both Raman modes display gradual changes with scan

direction. (f) The evolution of PL peak positions of alloyed monolayer $Mo_xW_{1-x}S_2$ as a function of distance along the scan direction. The gradual shift in the peak position along the scan direction indicates spatially varying optical band gaps within the material. (g) 3D cube hyperspectral PL map. (h) Section cut images taken from the hyperspectral map at different energies.

**Low-temperature photoluminescence of the alloyed monolayer $Mo_xW_{1-x}S_2$**

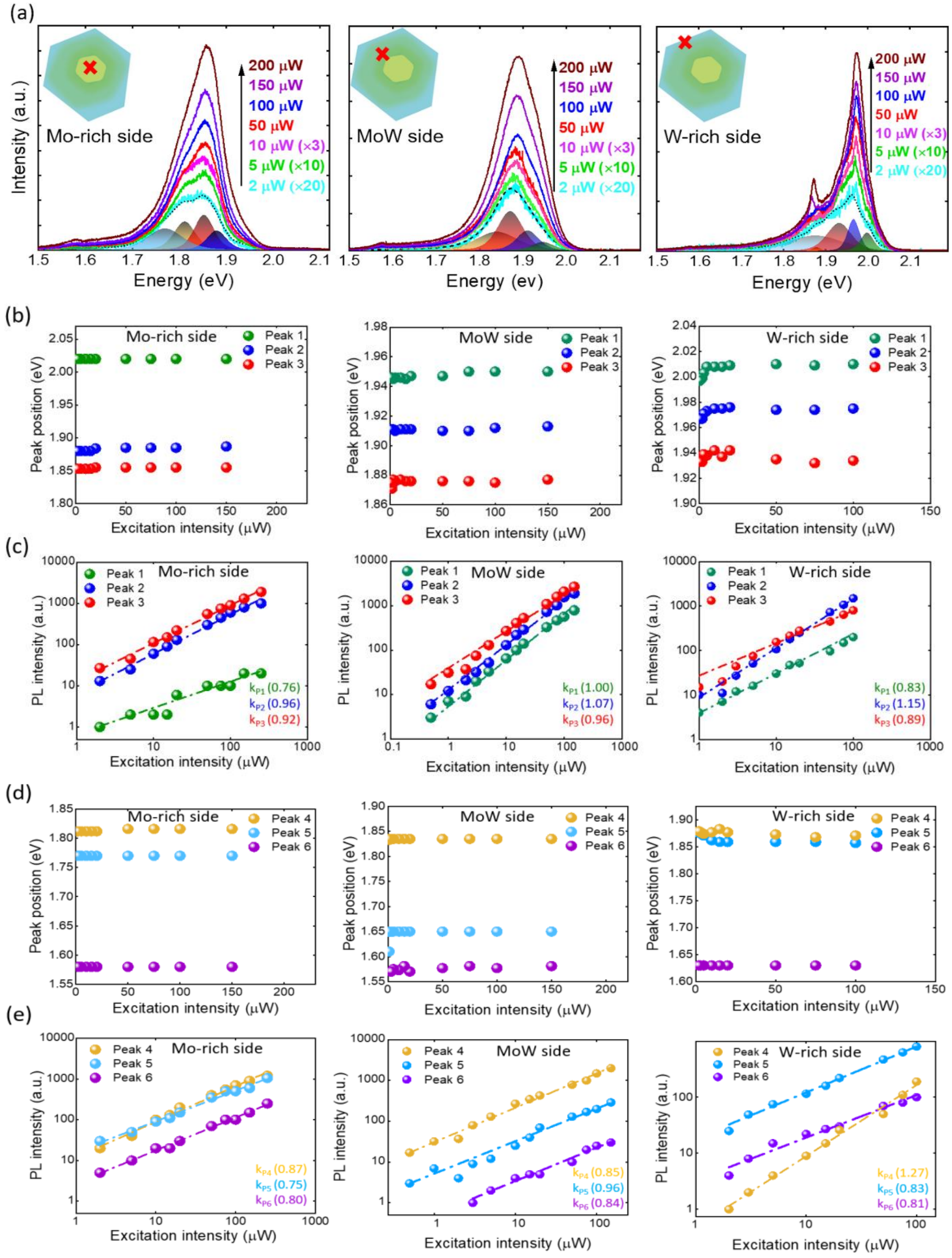

**Figure 3.** Laser excitation power dependence of the photoluminescence for the alloyed monolayer $Mo_xW_{1-x}S_2$ at T=4K. (a) shows the spectra from the center (Mo-rich side), intermediate region (MoW side), and edge (W-rich side). (b) Power dependence of the peak positions (peaks labeled 1-3) at the three regions. (c) PL intensity of peaks 1-3 at different

excitation powers at the three regions. (d) Power dependence of peak position for peaks 4-6 at the three regions. (e) PL intensity of peaks 4-6 at different excitation powers at the three regions.

The excitation power dependence of the PL intensities provide insight into the nature of the radiative recombination processes that give rise to the different spectral features near the band edge at different regions of the alloyed structure [36]. To that end we performed laser excitation power-dependent PL spectroscopy at low temperature (T=4 K). Figure 3a shows the PL spectra of the $Mo_xW_{1-x}S_2$ alloy with data obtained at the W-rich edge region (region 3), Mo-rich center region (region 1), and the intermediate composition region (region 2) acquired using excitation laser power from in the range 2 µW to 200 µW. Throughout the paper, the PL spectra for all powers and the three regions were deconvoluted using six pseudo-Voigt function peaks shown in the shaded colored representative spectra, which provide the best fit of the experimental data (Fig. 3a Peaks 2-5; peaks 1 & 6 are weaker compared to the other peaks and are not shown). Further representative spectra analysis is shown in Fig. SI1 at the edge, center, and the intermediate composition regions using excitation laser power of 100 µW at 4K.

**Near band edge peaks (1-3) in the three regions**

In order to study the recombination mechanisms of different peaks in different regions of the alloyed $Mo_xW_{1-x}S_2$ monolayer, we have divided the fitted peaks into two main categories depending on their location relative to the band edges. Firstly, we selected peaks close to the band edges (peaks1-3), then we chose peaks far away from those band edges (peaks 4-6). Figure 3b shows the peak positions for peaks 1-3 in three different regions of the alloyed monolayer as a function of excitation power to determine their origin. In the Mo-rich center region of the sample the highest energy peak (peak 1~ 2.02 eV) is weak for all the excitation powers. In this region, the position of peaks 1-3 show negligible change with excitation power. For the intermediate MoW region, peaks 1-3 show a blue shift of ~ 2 meV with increasing excitation power. This shift can be considered negligible as it is within the uncertainties of the spectral fitting. In the W-rich side, the peak shifts are larger, ~ 4 meV for peak 1 and ~ 9 meV for peak 2. Peak 3 shows initially a blue shift (~ 9 meV) and then a red shift (~ 4 meV). As the density of the photoexcited carriers increases with increasing excitation power, the quasi-Fermi levels for electrons and holes shift respectively into the conduction and valence bands leading to a blue shift [37,38]. However, other many-body interactions such as strain-induced band variation giving rise to reduced band gap and exciton binding energy would lead to a redshift of the excitonic peaks [39,40]. For the low excitation powers used in our experiments presumably the carrier density is not high enough to give rise to the many-body interactions and the observed blue shifts are essentially caused by the shift of the quasi-Fermi levels.

To determine the physical origin of the different peaks in different regions of the alloyed semiconductor, power dependence of the integrated of PL intensity of the various peaks are extracted. Figure 3c shows the PL intensity of peaks 1-3 as a function of excitation power using the power law $I=P^k$ fit where $I$, $P$, and $k$ represent PL intensity, laser excitation intensity, and a numeric coefficient, respectively. Typically, free and bound excitons show a linear dependence of excitation power (k=1), biexcitons show a quadratic dependence (k=2), and sub-linear (k<1) dependence would indicate free-to-bound type of radiative transitions through impurities and defects [41]. The k values are indicated for the different peaks in Fig 3c. A three-particle center like a trion would be expected to show a superlinear dependence (k=3/2). However, if the trions dominate the radiative recombination process, which is likely at high excitation intensities, then a linear dependence (k=1) is expected[42]. As evidenced in Fig. 3c, peaks 1-3 show nearly a linear dependence on excitation intensity in all the three regions. The low k value of 0.76 for peak 1 in Mo-rich side is attributed to the low PL intensity of the peak and the resultant uncertainties in the spectral fits. Based on the near linear dependence of the PL intensity on excitation intensity we surmise that the peaks 1-3 in all the regions are of excitonic origin.

By comparing the peak positions in Mo-rich side of the $Mo_xW_{1-x}S_2$ alloyed monolayer with the values reported for the A and B excitons in monolayer pristine $MoS_2$ (Table SI 2), we assign the peaks 1 (~ 2.01 eV) and 2 (~ 1.89 eV) in the Mo-rich side to the B and A excitons, respectively. Similarly, peak 1 in the W-rich side can be assigned to the A exciton. The B exciton in pristine $WS_2$ is generally much weaker than the A exciton and is rarely evidenced[43,44] in PL measurements. Accordingly, by comparing peak 1 in W-rich region (~ 2.01 eV) with the values reported for the A exciton in monolayer $WS_2$ (Table SI 1), we assign it to the A excitonic transition in the W-rich side. Thus, the difference in the peak positions assigned to the A excitons in Mo-rich side and W-rich side is 120 meV which is comparable to the 100 meV PL shift observed at room temperature from center (Mo-rich side) to edge (W-rich side) (Fig. 2c) in the $Mo_xW_{1-x}S_2$. As the band gap increases from the Mo-rich side to the W-rich side, the A excitonic position will increase continuously from the center to the edge region, we can assign peak 1 in region 2 (intermediate MoW side) to the A exciton. Thus, the A exciton peaks vary from 1.89 eV (Mo-rich side) to 1.95 eV (intermediate MoW side) and to 2.01 eV (W-rich side). Assuming similar exciton binding energies, this is consistent with the increase of the band gap going from $MoS_2$ to $WS_2$.

The near linear dependence of the PL intensities of the peak 3 (~ 1.86 eV, region 1), and peaks 2 (~1.91 eV, region 2) (~ 1.98 eV, region 3) as shown in Fig. 3c suggests they are of excitonic origin as well. Peak 3 (region 1) and peak 2 (regions 2 & 3) are separated from the respective A exciton peaks by 29 ± 2 meV in region 1, and 36 ± 2 meV and 33 ± 2 meV in regions 2 and 3, respectively. Generally, the PL peak observed on the low energy side of the A exciton in TMDCs is attributed to radiative recombination involving a three-particle (an exciton with an electron or a hole) trion [45,38] , with binding energies in the range 20-40 meV in pristine $MoS_2$ and 40-60 meV in pristine $WS_2$ (Tables SI 1 and SI 2). However, the expected k~ 3/2 dependence on excitation intensity of the assigned trion peak has not been reported[42]. Further, the calculated trion binding energies for $MoS_2$ and $WS_2$ are roughly the same, of the order of 30 meV, and are also very sensitive to the dielectric environment[46]. In an alloyed semiconductor such as $Mo_xW_{1-x}S_2$, the alloy disorder, if any, may also affect the trion binding energy. As shown in Fig. 3c, peak 3 (region 1) and peak 2 (regions 2 and 3) show k ~ 1 suggesting an excitonic recombination. A previous study in monolayer $WS_2$ assigned the peak on the low energy side, separated by ~ 29-35 meV from the A exciton, to a bound excitonic transition [33]. We conclude that peak 3 (region 1) and peak 2 (regions 2 & 3) are not trion-related transitions but are bound exciton transitions arising from the same impurity or defect in the alloyed $Mo_xW_{1-x}S_2$. Over the range of excitation powers used, peak 3 in region 2 (~ 1.88 eV) and region 3 (~ 1.93 eV) are separated by 71 ± 2 meV and 69 ± 5 meV, respectively, from the A exciton peak. This peak also shows linear dependence on excitation intensity implying a bound excitonic transition for its origin as well.

**Peaks (4-6) far from the band edge in the three regions**

As shown in Fig. 3a, the PL spectra from $Mo_xW_{1-x}S_2$ show significant broadening on the low energy side in all the three regions. The spectral fitting of this region yields three additional peaks, labeled P4, P5 and P6. The dependence of these peaks, observed far from the band edge, on excitation power is shown in Fig. 3d & 3e. The peaks generally show negligible shifts in their positions with increasing excitation power with the exception of peak 5 (~ 1.88 eV @ 1µW; 1.857 eV @ 100 µW) in W-rich side which shows initially a red shift of ~ 20 meV and saturates above 20 µW. It is expected in an alloy semiconductor that upon increasing the excitation intensity the photoexcited carriers migrate to regions of lower bandgap due to compositional grading before recombination. However, since the other peaks do not show such a large red shift, it is likely that peak 5 arises from a localized region of compositional disorder in region 3.

**DFT calculations**

To assist in the analysis and assignment of the PL transitions observed at low temperature, we performed DFT calculations to identify the role of alloying on the electronic band structure and in particular, on the energy levels of the predominant defect center, namely sulfur vacancy, $V_S$. Most theoretical work regarding the spin-orbit coupling effect in TMDCs considered pristine $MoS_2$ and $WS_2$, where the spin orbit splitting between the spin-up and spin-down states in $WS_2$ and $MoS_2$ monolayers are calculated to be 0.4 eV [47,48] and 0.15 eV [49-51] . This splitting effect is largely responsible for the production of the A and B excitons in these pristine samples [52,53] (observed in PL spectra measurements due to spin-allowed bright excitonic transitions). While a few studies have used spin-orbit coupling for $WS_2$ and $MoS_2$ monolayers, there are not many studies involving the spin-orbit coupling effect with the $V_S$ defect. When the $V_S$ defect is considered, calculations reveal that the valence band splitting tends to decrease to 0.05 eV and 0.30 eV for both $MoS_2$[54,55] and $WS_2$[56], respectively. STM imaging of the $V_S$ defect has confirmed the valence band splitting for $WS_2$ to be 0.25 eV [56] while only theoretical calculations have been shown to yield a decrease in valence band splitting for $MoS_2$ with the $V_S$ defect. The $V_S$ defect introduces defect levels within the bandgap, which are radiative and can be attributed to the origin of PL peaks below the A exciton [33,57]. Given that spin-orbit coupling has not been studied in $Mo_xW_{1-x}S_2$ alloys with varying Mo concentration, this study attempts to both understand the evolution of the A and B exciton and the $V_S$ defect-mediated radiative transitions by investigating multiple possible configurations for the $Mo_xW_{1-x}S_2$ alloy and examine how by varying the positions and concentration of W and Mo atoms surrounding the $V_S$ defect affects the optical transitions.

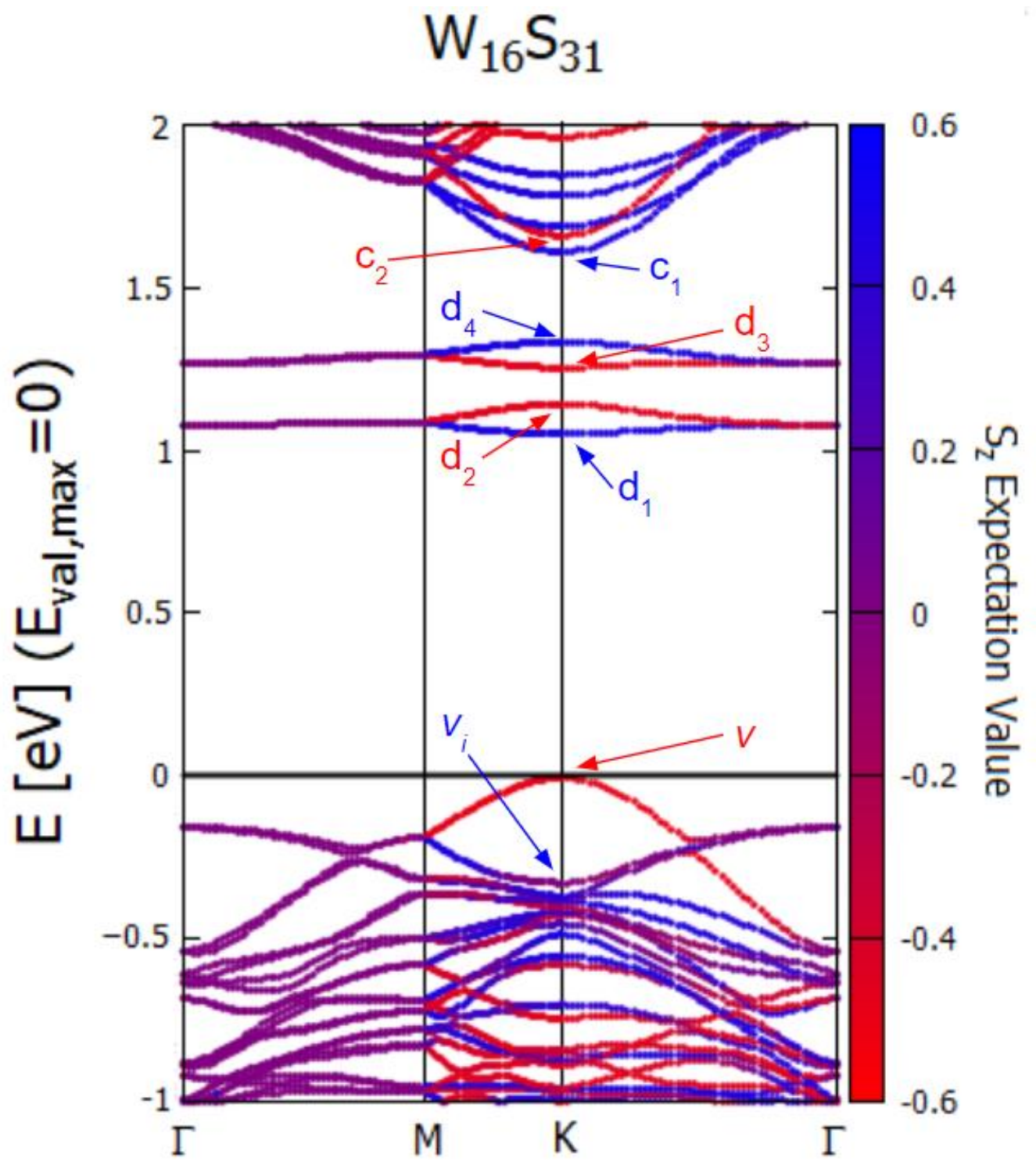

**Figure 4.** The $W_{16}S_{31}$ band structure calculated using the PBE functional to illustrate the energy levels considered in the alloys. Note the spin-orbit split doublet structure of the $V_S$ level in the gap.

### Sulfur vacancy, $V_S$ levels (Computational Methods)

The energy levels are all considered at the K point with the distinct energy levels being denoted in the valence band, defect band, and conduction band (Fig. 4) with $W_{16}S_{31}$ as an example. In all cases investigated, only the allowed transitions are considered. That is, the energy difference between bands of the same color (blue to blue and red to red) are allowed. The $v$-$c_2$ and $v_i$-$c_1$ transitions in the pristine alloys [$Mo_{16x}W_{16(1-x)}S_2$] are attributed to the A and B excitons, respectively. The defect-mediated transitions ($v$-$d_2$, $v$-$d_3$, $v_i$-$d_1$, and $v_i$-$d_4$) are obtained from the $V_S$ hosting alloys [$Mo_{16x}W_{16(1-x)}S_2$] and are studied to develop a model to predict the transition energies based on the geometry of the alloy. In addition, the valence ($v_i$-$v$) and conduction band splitting ($c_1$-$c_2$) are also studied. For pristine $W_{16}S_{32}$ and $Mo_{16}S_{32}$, the A (B) excitonic transition energies are underestimated and are measured to be 1.607 eV (2.001 eV) and 1.615 eV (1.759 eV), respectively, with conduction (valence) band splitting of 0.032 eV (0.426 eV) and 0.003 eV (0.147 eV), respectively. Given the transitions are underestimated, a scissor shift is performed to correct the A (B) exciton energies to the experimental energies of 1.88 eV (2.02 eV) and 2.02 (2.40 eV) for $Mo_{16}S_{32}$ and $W_{16}S_{32}$, respectively [58,59].

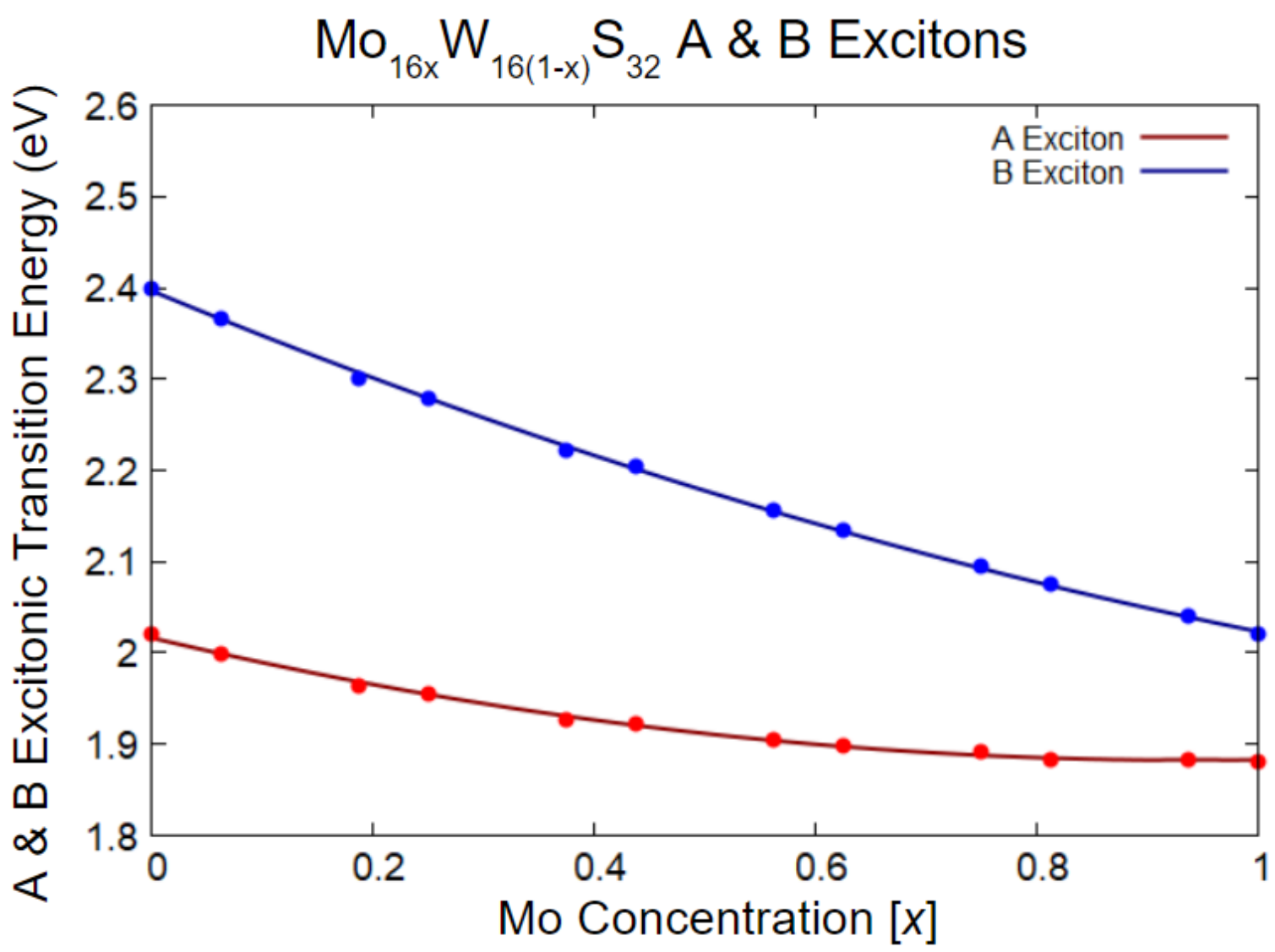

**Figure 5.** The evolution of the spin-down (A-Exciton) and spin-up (B-Exciton) versus Mo concentration [$x$] in pristine [$Mo_{16x}W_{16(1-x)}S_{32}$] alloys where the transition energies are shifted from the PBE functional exciton energies to the experimental exciton energies.

The evolution of the A exciton tends to decrease monotonically with the decrease in energy becoming less rapid as the Mo concentration [$x$] increases (Fig. 5), which confirms with the observations seen in the experimental PL spectra measurements (Fig. 2, Tables SI 1-2). The 32 configurations investigated in this study are divided into three separate classes based on the Mo concentration [$x$]: Mo-rich ($x \geq 0.75$), MoW [intermediate region] ($0.25 < x < 0.75$), and the W-rich ($x \leq 0.25$) regions. For each of these regions, the range of transition energies for each of the defect-mediated transition energies ($R_1$, $R_2$, $B_1$, and $B_2$ as shown in Fig. SI 19) for each of these three classes is shown in Table SI 4. In addition, since the defect levels tend to be significantly closer to the conduction band generating shallow defect levels when the $V_S$ defect is surrounded by W atoms, a separate range is shown by separating the range of the transition energies obtained when the $V_S$ defect is surrounded by Mo and W atoms to emphasize the possibility of $V_S$ defects surrounded by W atoms contributing to the PL peaks below the A exciton [60,61] An example of this effect with the $Mo_{10}W_6S_{31}$ configurations is shown in Figure SI 21 in the supplementary data.

When the $V_S$ defect is surrounded by W atoms, the ranges for the defect-mediated transitions match more closely to the P5 and P6 PL peaks observed in the experiment. Defect-mediated transitions associated with the $d_2$ and $d_3$ defect levels are assigned to the P5 and P6 PL peaks across the three distinct regions in Table 1 and Table SI 4. From the excitation intensity of the PL intensities of P4-P6 shown in Fig. 3e it can be seen that most of the peaks show a sublinear dependence ($k < 1$) on excitation intensity with the exception of P4 showing a superlinear ($k > 1$) dependence in region 3 (W-rich side) and P5 showing a linear dependence $k \sim 1$) in region 2 (MoW side). While the peaks do not show a clear $k \sim ½$ dependence, the sublinear dependence may suggest their origin as due to free-to-bound type transitions. It is likely that in the spectral region away from the band edge the peaks may arise from both bound excitons and/or free-to-bound transitions arising from impurities/defects having energy levels within the gap of the alloyed semiconductor. Since the peaks away from the band edge should also follow the band gap increase from the Mo-rich side to the W-rich side of the $Mo_xW_{1-x}S_2$ alloyed monolayer, we can assign peak 4 (~ 1.82 eV) in region 1 and peak 3 in region 2 and 3 (~ 1.877 eV in region 2, 1.934 eV in region 3) to the same origin. The shift in the peak positions from region 1 to region 3 is ~ 114 meV. Similarly, P4 in region 2 and 3 (~ 1.835 eV in region 2, 1.871 eV in region 3) and P5 in region 3 (1.857 eV) are conjectured to be of the same origin. P6 (~ 1.6 eV) seen in all the three regions appears to be band gap independent. It should be noted that the P4-P6 peaks are not observed in the low temperature spectra from pristine monolayer $WS_2$ (see Fig. SI 13). Therefore, these features can be assumed to be caused by the presence of Mo in the $Mo_xW_{1-x}S_2$ alloyed monolayer.

Thus, based on the discussion above and first principles calculation of the energy levels of the most likely point defect in alloyed $Mo_xW_{1-x}S_2$, namely, $V_S$, the following assignments of peaks in the different regions can be made as shown in Table 1. The various radiative recombination paths of peaks 1-6 in the three regions are illustrated in Fig. 6. In the Mo-rich side peak 1 marked as B-exciton ($E_B$). The peaks identified with bound excitons in all the three different regions of the alloyed $Mo_xW_{1-x}S_2$ monolayer are shown with different binding energies ($\Delta$) which are defined as the energy separation of the peak from the A-exciton peak ($E_A$). The free-to-bound transitions associated with $V_S$ level, $d_3$ and $d_2$ as well with the unknown defect X are also shown in Fig. 6(a-c). The excitonic positions in monolayer of pristine $MoS_2$ and $WS_2$ are summarized in Tables SI 1-SI 2.

Table 1. Peak assignments for the alloyed monolayer $Mo_xW_{1-x}S_2$, pristine $MoS_2$, and pristine $WS_2$ at T=4K.

| **Peak position assignment** | Pristine $MoS_2$ (eV) | Pristine $WS_2$ (eV) | Mo-rich side of alloyed (eV) (Region 1) | MoW side of alloyed (eV) (Region 2) | W-rich side of alloyed (eV) (Region 3) |
|---|---|---|---|---|---|
| B-exciton | 2.05 | - | 2.01 (P1) | - | - |
| A-exciton($X^0$) | 1.895 | 2.04 | 1.89 (P2) | 1.95 (P1) | 2.01 (P1) |
| Bound exciton $\boldsymbol{V_S^{BE}(d_3)}$(associated with $V_S$ level, $d_3$, Fig. 4) w/ exciton binding energy ~ 20-36 meV[1] | 1.875 | 2.01 | 1.86 (P3) | 1.913 (P2) | 1.975 (P2) |
| Bound exciton transition $\boldsymbol{V_S^{BE}(d_2)}$ (associated with $V_S$ level, $d_2$, Fig. 4) w/ exciton binding energy ~ ~ 70 meV[1] | - | 1.95 | 1.82 (P4) | 1.88 (P3) | 1.93 (P3) |
| Bound exciton/free-to-bound transition (unknown defect) ($\mathbf{X^{BE}}$)/($\mathbf{X^{FB}}$) w/ exciton binding energy ~ ~132 ± 7 meV[1] | - | - | - | 1.835 (P4) | 1.871 (P4) 1.857 (P5) |
| Free-to-bound transition $\boldsymbol{V_S^{FB}(d_3)}$ (associated with $V_S$ level, $d_3$, Fig. SI 19) | 1.73 | - | 1.77 (P5, R2) | 1.65 (P5, R2) | 1.63 (P6, R2) |
| Free-to-bound transition $\boldsymbol{V_S^{FB}(d_2)}$ (associated with $V_S$ level, $d_2$, Fig. SI 19) | 1.68 | - | 1.58 (P6, R1) | 1.58 (P6, R1) | - |

[1] Binding energy is defined as the energy separation of the peak from the A-exciton peak [33].

A broad band observed at ~ 1.75 eV in monolayer $MoS_2$ similar to the P5 band in Mo-rich region of the alloyed $Mo_xW_{1-x}S_2$ monolayer in our study has been identified with an exciton bound to ionized donor levels, related to $V_S$ [62]. We assign the bands P5 and P6 in region 1 to a free-to-bound transition between the photoexcited electron captured at the $V_S$ levels, $d_3$ and $d_2$, respectively and a hole in the valence band. On the other hand, the peaks P3 and P4 in Mo-rich region are identified with recombination through excitons bound to $V_S$ levels, $d_3$ and $d_2$, respectively. The peaks P4 in region 2 (MoW side) and P4 & P5 in W-rich region do not correspond to the calculated energy level positions of $V_S$ and presumably arise from an unknown defect/impurity. It should be noted that P4 in W-rich side is considerably sharper than P5. Further P4 shows a superlinear dependence on excitation intensity ($k > 1$) in region 3 while P5 has a sublinear ($k < 1$) dependence. It may be conjectured that in region 3, P4 and P5 arise from the same defect/impurity with the former being a bound exciton and the latter being a free-to-bound transition. In intermediate MoW region, P4 may be a free-to-bound transition associated with the same defect/impurity. This defect/impurity is introduced as the W content increases.

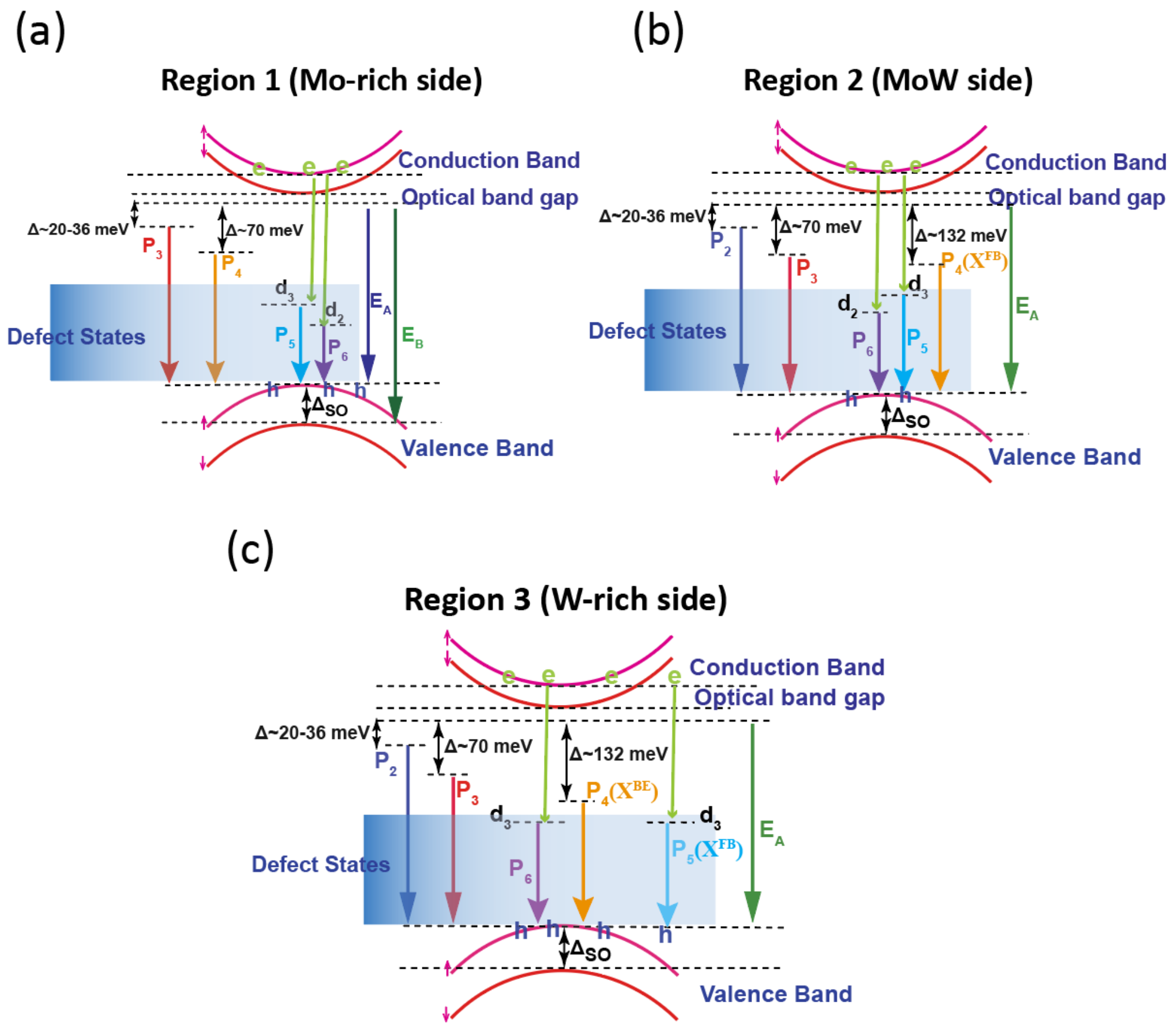

**Figure 6.** Various radiative recombination paths of peaks 1-6 at three regions of the alloyed $Mo_xW_{1-x}S_2$ monolayer. The energy band diagram is shown for (a) Mo-rich side, (b) intermediate

MoW side, and (c) W-rich side of the alloyed $Mo_xW_{1-x}S_2$ monolayer. The bound exciton transitions through the doublet $V_S$ levels ($d_3$, $d_2$) are indicated as peaks P3 & P4 in Region 1, and P2 & P3 in Regions 2 and 3. The free-to-bound transitions through $d_3$ and $d_2$ are indicated by P5 and P6 in Region 1 and Region 2. The free-to-bound transition through $d_3$ in Region 3 is indicated by P6. The bound exciton through the unknown defect X is indicated by P4 in Region 3. The free-to-bound transition through the defect X is indicated by P4 in Region 2 and by P5 in Region 3.

**Temperature dependence of PL in alloyed $Mo_XW_{1-X}S_2$**

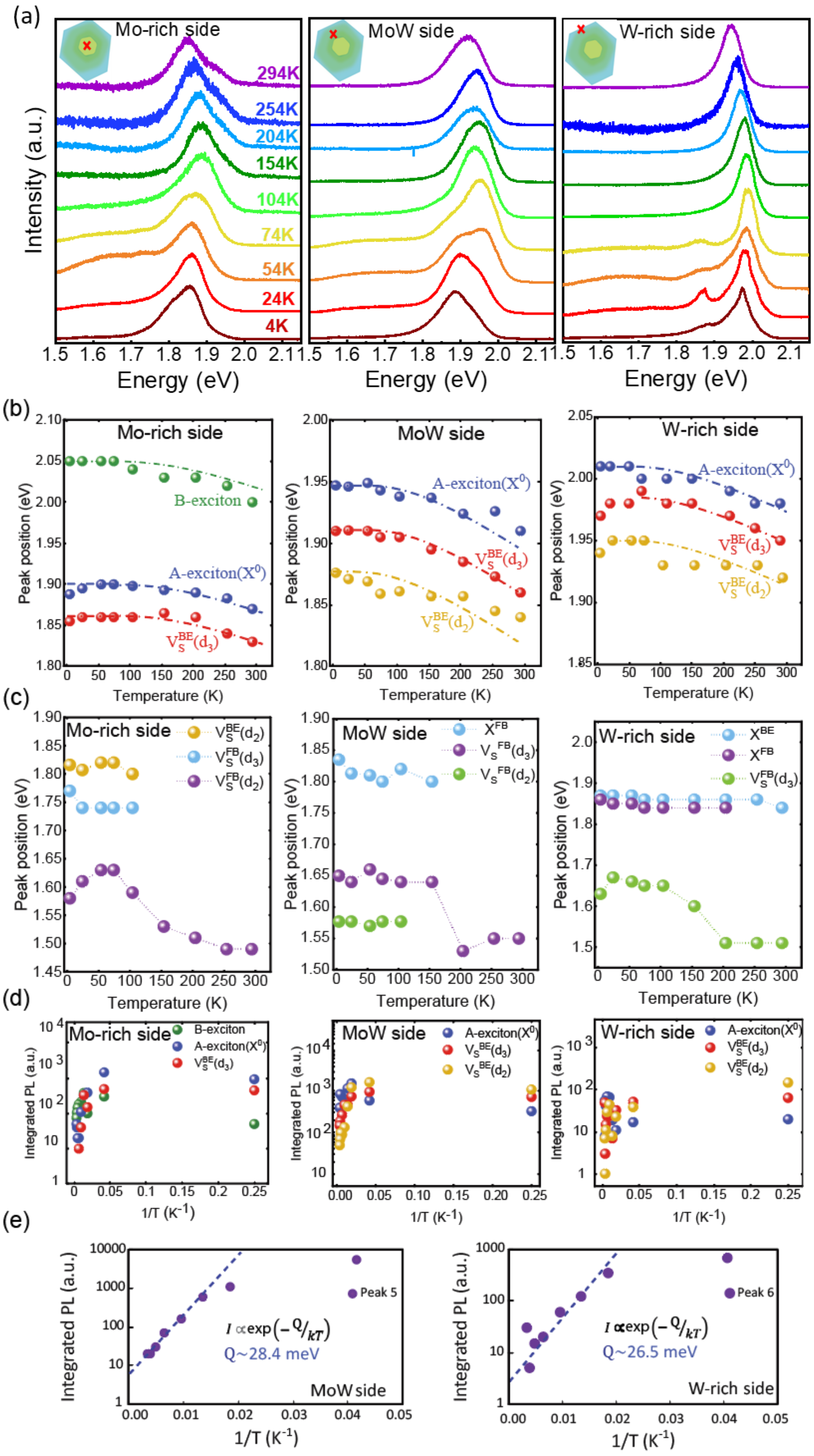

**Figure 7.** Evolution of PL Spectra of alloyed $Mo_xW_{1-x}S_2$ monolayer. (a) shows PL Spectra as a function of temperature for alloyed structure in region1 (Mo-rich side), region2 (intermediate MoW side), and region3 (W-rich side) of the flake. (b) Evolution of peak position with temperature for 6 peaks at the three different regions of alloyed monolayer $Mo_xW_{1-x}S_2$. Dash-dot lines are calculated with the conventional temperature dependence of semiconductor bandgap. (d) Temperature dependence of the PL intensity for the near band edge peaks for alloyed $Mo_xW_{1-x}S_2$ monolayer. (e) Temperature dependence of the PL intensity for peaks P5 and P6 from regions 2 (Intermediate MoW side) and 3 (W-rich side), respectively. Fitting an Arrhenius relation to the data yields an activation energy of 28.4 meV (P5) and 26.5 meV (P6).

To further investigate the nature of excitons and the effect of the alloy in $Mo_xW_{1-x}S_2$, temperature dependent PL spectra were taken at different regions. Figure 7(a) depicts the evolution of PL spectra for three different regions of the alloyed $Mo_xW_{1-x}S_2$ monolayer in the temperature range 4 K–300 K. (A 532 nm excitation laser was used at fixed power of 50 µW, and laser exposure time was kept at 2 sec for all spectra.) With increasing temperature, optical transition energies (peak positions) show a red shift as expected from the temperature dependence of the bandgap. The temperature dependence of the different peaks in the three regions of alloyed $Mo_xW_{1-x}S_2$ are shown in Fig. 7(b) and Fig. 7(c). We applied the O'Donnell equation[63] to fit the experimental peak positions near the band edge (peaks1-3) at different temperatures (dot dash-lines in Fig. 7b) for all three regions of the alloyed monolayer $Mo_xW_{1-x}S_2$ as given by:

$$E_g(T) = E_g(0) - S < \hbar\omega > \left[coth\left(\frac{<\hbar\omega>}{2kT}\right) - 1\right] \quad (2)$$

where $E_g(0)$ is the ground-state transition energy at 0K, S is a dimensionless coupling constant and $<\hbar\omega>$ is an average phonon energy, respectively. Table SI 3 shows the fitting parameters of peak positions at the different regions of the alloyed monolayer $Mo_xW_{1-x}S_2$. Based on the fitting the average phonon energy <ℏω> for A-exciton in pristine $MoS_2$, and in Mo-rich side of the alloyed monolayer $Mo_xW_{1-x}S_2$ is 50 meV which is comparable to the reported values [64,65] . In addition, the average phonon energy <ℏω> for A-exciton ($X^0$) in both pristine $WS_2$, and W-rich side of the alloyed monolayer $Mo_xW_{1-x}S_2$ is ~35-44 meV which is comparable to the reported values [66,67]. Based on the Fig. 7(b) the red-shift in the peak position of B-exciton in the Mo-rich side is ~ 50 meV when the temperature increases from 4K to 300K which is comparable with reported values [68]. The red shift for A-exciton in all the three regions is about (~33±3 meV) when the temperature increases from 4K to 300K.

The peaks assigned to bound excitons associated with the $V_S$ levels $d_3$ and $d_2$ (see Table 1) are generally stable up to room temperature with the exception of the bound exciton associated with the $d_2$ level in region 1 (Mo-rich side) of the alloy which disappears above 100K (Fig.7c). The peaks assigned to an unknown defect with a large exciton binding energy (132 ± 7 meV, Table 1) in regions 2 (intermediate region) and region 3 (W-rich side) persist at high temperatures, especially in W-rich region of the alloyed $Mo_xW_{1-x}S_2$ monolayer (Fig.7c). The radiative transitions assigned to free-to-bound transitions involving $V_S$ levels $d_3$ ($V_S^{FB}(d_3)$) are present even at room temperature. (See Fig. SI 8 for a comparison of the PL spectra at low and high temperatures from the three regions). Interestingly, $V_S^{FB}(d_3)$ and $V_S^{FB}(d_2)$ transitions, are very prominent at low temperature in pristine $MoS_2$ (see Fig. SI 12) and persist as a broad tail on the low energy side of the band excitons even at high temperatures. This observation suggests that the $V_S$ levels dominate the radiative process in pristine $MoS_2$ especially at low temperatures. While the free-to-bound transitions via $V_S$ levels dominate in pristine $MoS_2$, the bound excitons associated with $V_S$ levels are observed in pristine $WS_2$ in the temperature range 4-304 K (see Fig. SI 13). Thus, $V_S$ plays an important role in the radiative processes in both $MoS_2$ and $WS_2$ as well as in the alloyed monolayer $Mo_XW_{1-X}S_2$. The dominance of either the free-to-bound transitions or the bound exciton transitions involving the $V_S$ levels will be determined by the relative radiative rates of the transitions and the thermal ionization of the photoexcited electron from the $V_S$ level, especially at high temperatures, which may differ between $MoS_2$ and $WS_2$ layers.

Figure 7(d) shows the integrated PL intensity versus reciprocal of temperature for near the band edge (peaks 1-3) in all the three regions of alloyed monolayer $Mo_xW_{1-x}S_2$. The PL intensity ($I_{PL}$) decreases at high temperatures suggesting an Arrhenius type behavior[69, 10]. The quenching of the free-to-bound transitions via the $d_3$ level of $V_S$ (P5 in region 2 and P6 in region 3, Table 1) is shown Fig. 7(e). Fitting an Arrhenius relation $I_{PL} = const * e^{\frac{Q}{KT}}$ to the data gives an activation energy Q ~ 26-28 meV. This value perhaps suggests the thermalization energy of the photoexcited electron captured at the $d_3$ level of $V_S$. A comparable activation energy of ~ 36 meV was reported for PL emission associated with $V_S$ in monolayer $WS_2$ and was attributed to the thermal dissociation of the bound exciton [33].

Figure SI 3 shows the FWHM versus temperature for the near band edge peaks (peaks 1-3) in the three regions of the alloyed $Mo_xW_{1-x}S_2$. It can be seen that the temperature dependence of FWHM could not be fitted to the equation describing the electron-phonon interaction (EQ (1) in SI). The A-exciton peak is broader in regions 1 and 2 in comparison to region 3. The FWHM of the peaks associated with bound exciton transitions involving $V_S$ levels are in the range 60-80 meV. Compared to the pristine $WS_2$ and $MoS_2$, the changes of the FWHM for A-exciton ($X^0$) in region 1 and region 2 of the monolayer $Mo_xW_{1-x}S_2$ is larger than the changes of FWHM for pristine samples. For region 1 (Mo-rich side) of the alloy it changes from 70 meV to 120 meV upon

increasing the temperature from 4 K to 300 K. However, for the pristine $MoS_2$ FWHM of the A-exciton changes from 38 meV to 50 meV (Fig. SI 15). For the region 3 (W-rich side) of the alloy it changes from 38 meV to 45 meV upon increasing the temperature from 4 K to 300 K (Fig. SI 3). For pristine $WS_2$, the FWHM changes from 45 meV to 52meV (Fig. SI15) which is larger than the reported values for pristine $MoS_2$[70], and pristine $WS_2$ [67].

## CONCLUSION

In summary, using an alkali metal halide-assisted chemical vapor deposition approach, we successfully synthesized triangular TMDC alloy ($Mo_xW_{1-x}S_2$) monolayers with continuously varied W/Mo concentration from the center to edges. The combined experimental photoluminescence characterization with theoretical DFT calculations including spin-orbit interactions enabled thorough investigation of the nature of myriad intralayer optical transitions excitons in the alloyed monolayers. Aberration-corrected high-angle annular dark-field scanning transmission electron microscopy showed the presence of sulfur monovacancy, $V_S$, whose concentration varied across the graded $Mo_xW_{1-x}S_2$ layer as a function of Mo content with the highest value in the Mo rich center region. We identified free-to-bound transitions involving a photoexcited electron captured at the doublet $V_S$ level and a hole in the top of the valence band by matching the calculated spin-allowed optical transition energies through a doublet $V_S$ level in the gap. In addition, two bound exciton transitions associated with the $V_S$ doublet were also identified. Further, the study of the temperature dependence of the photoluminescence helped to identify the differences brought about by alloying in comparison to the pure $MoS_2$ and $WS_2$ monolayers. Thus, a plethora of $V_S$ related intralayer optical transitions reported for the first time in the alloy $Mo_xW_{1-x}S_2$, reveals the interplay between composition and defect structure. Our work highlights the capability of modulating the bandgap and engineering the defect structures simultaneously in 2D graded TMDC alloys via controllable synthesis strategies, and systematically studies how the optical transitions are modulated by the presence of structural defects.

## MATERIALS AND METHODS

**The Growth of Monolayer $Mo_xW_{1-x}S_2$ alloys.** To grow alloyed $Mo_xW_{1-x}S_2$ monolayers, powders of $MoS_2$ (~5 mg), $WO_3$ (~5 mg), and NaBr (~0.5 mg) were mixed uniformly and placed inside a porcelain boat, and a piece of clean $SiO_2$/Si was placed on top of the boat with the polished side facing down. Note that the use of NaBr promoter can enhance the coverage of monolayers and increase the reproducibility of alloys. Subsequently, this porcelain boat was loaded into a one-inch quartz tube for the CVD growth, and another porcelain boat containing sulfur powders (300 mg) was loaded upstream. In the growth process, the mixed $MoS_2$/$WO_3$/NaBr powders and the substrate were heated up to 825 °C and held for 10 min, and sulfur powders were heated up to 220 °C simultaneously for evaporation. Argon gas (100 sccm) was used as the carrier gas throughout the growth process.

**AC-HAADF-STEM.** High-resolution aberration controlled HAADF-STEM images were taken using a FEI Titan3 G2 S/TEM operated at 80 kV and equipped with double spherical aberration correction and monochromator. Images were acquired using a HAADF detector with a collection angle of 42-244 mrad, camera height of 115 mm, convergence angle of 30 rad, and a beam current of 50 pA. For the identification of sulfur mono-vacancies, AC-HRSTEM images were Fourier filtered using a low pass filter in the Digital Micrograph suite. In the same software, integrated intensity line scans were taken along the armchair direction and the relative intensity of two overlapped sulfur atoms was used as a reference to identify sulfur mono-vacancies.

**Optical Measurements.** The power and temperature-dependent PL spectra were measured by the confocal laser scanning microscope system equipped with a vibration-free closed-cycle cryostat (Attodry 800, attocube). A 532 nm CW laser as an excitation source was focused into a small spot with a diameter of approximately 2-3 μm on the sample through a 100× objective lens (APO/VIS, N.A. = 0.82; attocube) inside the vacuum chamber. The PL spectra was then collected by the same lens and filtered the excitation signal by a 532 nm long-pass filter before entering a spectrometer (Andor) which consisted of a monochromator and a thermoelectrically cooled CCD camera. Room temperature hyperspectral PL was collected using AFM assisted diffraction limited PL (neaspec co.) recorded using a 328 mm focal length Andor spectrometer and imaged with a liquid nitrogen cooled silicon EMCCD camera (Andor iXon).

**DFT Calculation.** We have performed theoretical calculations using Quantum Espresso (QE) [71] to aid in the interpretation of experimental results. The alloys are modeled using a hexagonal unit cell consisting of 3 atoms (either $MoS_2$ or $WS_2$) and are expanded to create a 4 x 4 x 1 supercell consisting of 48 atoms with stoichiometry $Mo_{16x}W_{16(1-x)}S_{32}$ and varying Mo concentration ($x$). All calculations are performed using norm-conserving PBE [72] pseudopotentials [73] with the spin-orbit coupling (SOC) interaction included [74]. The lattice constant is set to $a$ = (3.183 - 0.002$x$) Å, which is based on the relaxed lattice constants of 3.181 Å for $MoS_2$ and 3.183 Å for $WS_2$. The force convergence threshold is set to 0.01 eV/Å, while the total energy threshold is set to $10^{-6}$ eV. Also, calculations are performed using a 2 x 2 x 1 k-point mesh and the kinetic energy cutoff is 680 eV. The cell parameter in the z-direction is set to $c$ = 14.2 Å to introduce a vacuum region and prevent undesired interactions between the monolayers. All possible alloy structures are constructed by substituting Mo and W atoms at the transition metal sites while preserving the $C_{3v}$ symmetry. Assuming symmetry preservation, 32 possible alloyed structures can be generated in the 4 x 4 x 1 supercell and the band structures for all 32 structures are determined both with and without a single $V_S$ introduced (Fig. SI 18). All band structures are calculated using a Γ-M-K-Γ k-point path with the z-directional electronic spinor $\langle S_z \rangle$ also calculated to distinguish between spin up (blue) and down (red) states. From the calculated band structures, the various transitions at the K point are investigated to determine how the Mo concentration and geometry of the configuration affect these transitions.

**ACKNOWLEDGMENTS**

M.G. and Y.A. acknowledge support from Air Force Office of Scientific Research (AFOSR) grant number FA9550-19-0252 and FA9550-23-1-0375. T.Z., D.Z., D.H. and M.T. acknowledge support from the AFOSR through grant No. FA9550-18-1-0072 and the NSF-IUCRC Center for Atomically Thin Multifunctional Coatings (ATOMIC). H.T. and Z.D.W. acknowledge funding support from the USA National Science Foundation (Award 2013640) and are grateful to the "Advanced Cyberinfrastructure Coordination Ecosystem: Services & Support" (ACCESS), which is supported by National Science Foundation grant number ACI-1548562 through proposal TG-DMR170008.

**Author Contribution**

Y.A. and M.G. conceived the project. M.G. carried out the power and temperature-dependent PL experiments. T.Z., D.Z, and D.S. grew the samples and AC-HRSTEM measurements. M.G., and V.S. analyzed the experimental data. Z.D.W, and H.T. performed the DFT calculations. All authors discussed the results and contributed to writing the manuscript. Y.A. supervised the overall project.

## Competing interests

The authors declare no competing interests.

## Materials & Correspondence

All correspondence and material requests should be addressed to yohannes.abate@uga.edu.

# Supplementary data

**Sulfur Vacancy Related Optical Transitions in Graded Alloys of $Mo_xW_{1-x}S_2$ Monolayers**

Mahdi Ghafariasl [1], Tianyi Zhang [2], Zachary D. Ward [3], Da Zhou [4], David Sanchez [2], Venkataraman Swaminathan [5], Humberto Terrones [3], Mauricio Terrones [2,4,6], Yohannes Abate [1*]

[1] Department of Physics and Astronomy, University of Georgia, Athens, Georgia 30602, USA

[2] Department of Materials Science and Engineering, The Pennsylvania State University, University Park, PA 16802, USA

[3] Department of Physics and Astronomy, Rensselaer Polytechnic Institute, Rensselaer, New York 12180, USA

[4] Department of Physics, The Pennsylvania State University, University Park, PA 16802, USA

[5] Department of Materials Science and Nanoengineering Rice University Houston, TX 77005, USA

[6] Department of Chemistry, The Pennsylvania State University, University Park, PA 16802, USA

*Corresponding author E-mail: yohannes.abate@uga.edu

**PL spectra at 4K and fitting parameters at different regions of alloyed $Mo_xW_{1-x}S_2$ monolayer**

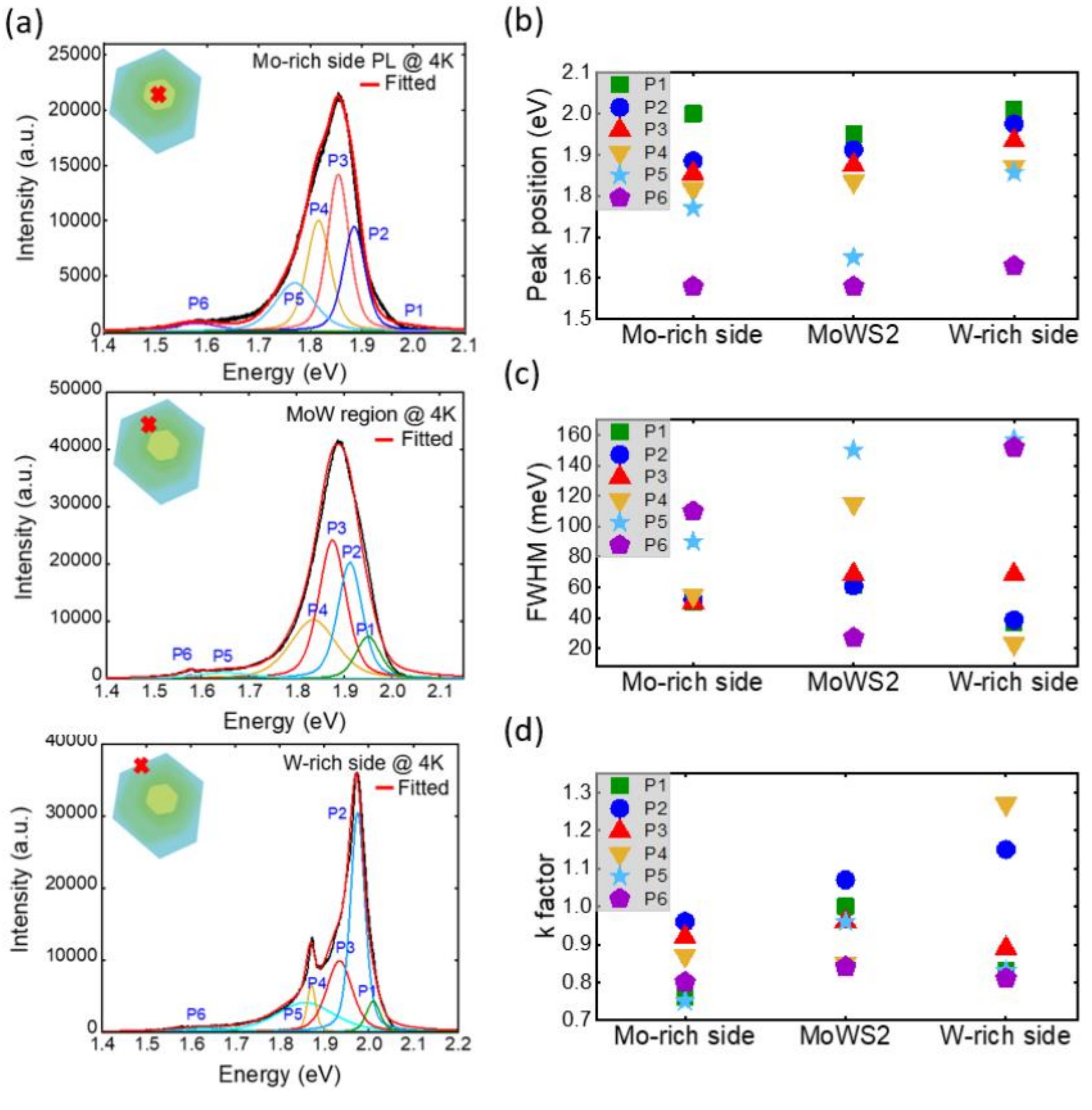

**Figure SI 1**. PL spectra at 4K and fitting parameters at different regions of alloyed monolayer $Mo_xW_{1-x}S_2$. (a) Peaks P1-P6 represent the pseudo-Voigt fits in three different regions of the alloyed monolayer $Mo_xW_{1-x}S_2$; Mo-rich region (center), an intermediate region, and W-rich region (edge), respectively. (b) Comparison of the peak positions in the different regions. (c) Comparison of FWHM of the different spectral peaks in the three regions. (d) Comparison of the k factor from the excitation dependence of PL intensity of the different spectral peaks in the various regions. See the main article for the discussion on the k factor. For PL spectra the power is 100 µW and the temperature of 4K.

The comparison of the A-exciton peak positions in the Mo-rich region (P2), the intermediate region (P1), and the W-rich region (P1), shows that the band edge PL peaks shift monotonically to higher energy from center to edge consistent with the center being Mo-rich and the edge being W-rich (Fig. SI 1 (b)). Figure SI 1(c) shows the comparison of the FWHM of different spectral peaks (P1-P6) in different regions of the alloyed structure. In general, peaks P4-P6 are broader (≥ 100 meV) than the band edge transitions P1-P3 (≤ 60 meV) and the latter are narrower at the edge (W-rich) than at the center (Mo-rich). Peak 4 on the W-rich side which is attributed to a bound exciton due to an unknown defect/impurity shows the narrowest linewidth. (See Table 1 in the main article for the peak assignments). The k-factors for the excitation dependence of PL intensity of the different spectral peaks are plotted in Fig. SI 1(d) and the assignments of the peaks based on their values are discussed in the main article.

**Power dependence of FWHM for alloyed monolayer $Mo_xW_{1-x}S_2$**

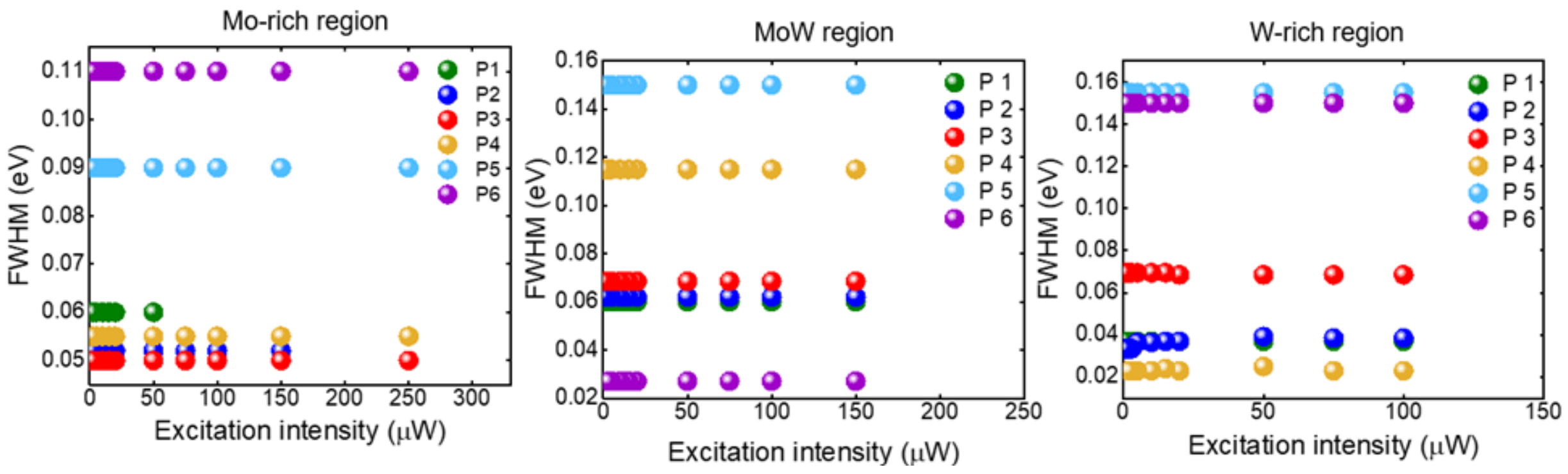

**Figure SI 2**. Power dependence of FHWM for the six peaks at three different regions of alloyed monolayer $Mo_xW_{1-x}S_2$ showing negligible effect.

**Temperature dependence of FWHM for alloyed monolayer $Mo_xW_{1-x}S_2$**

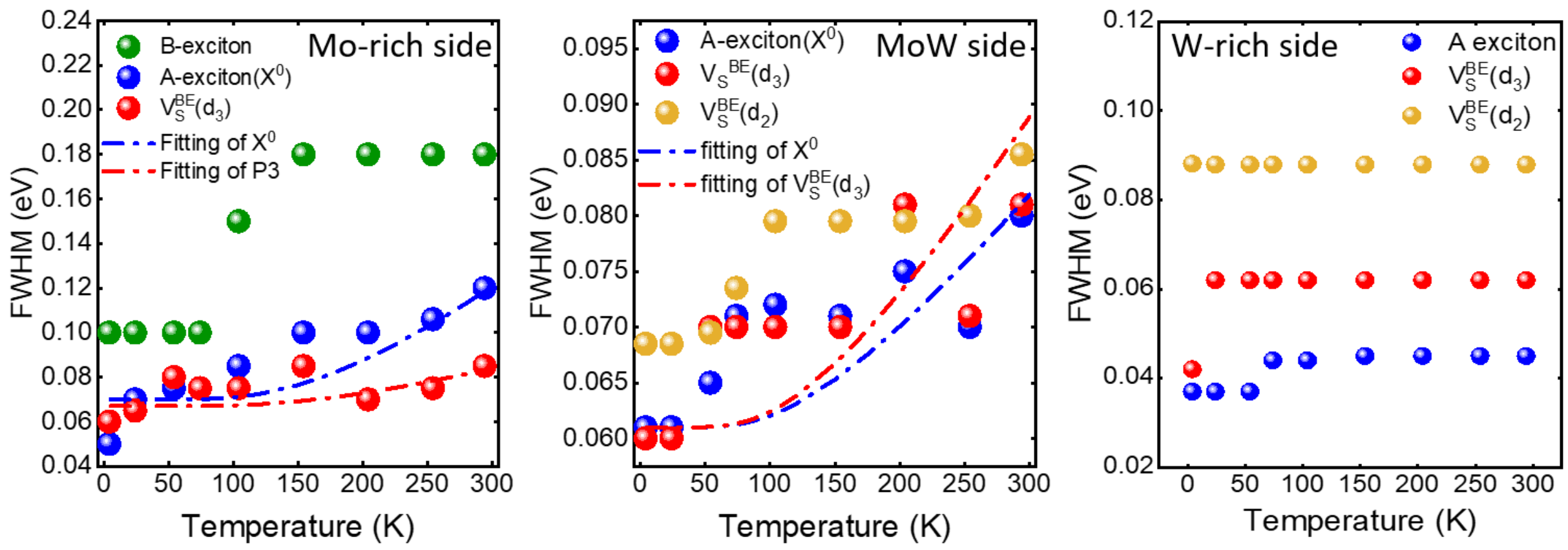

**Figure SI 3**. Temperature dependence of FWHM for alloyed monolayer $Mo_xW_{1-x}S_2$

The linewidth of the exciton transition as a function of temperature in TMDCs has been explained in terms electron-phonon interactions as given by the following equation [1,2]

$$\gamma = \gamma_I + \frac{b}{exp\left(\frac{\Theta}{KT}\right)-1} \qquad (1)$$

Where $\gamma_I$ is the temperature independent inhomogeneous broadening, Θ is either a dominant phonon or an average phonon energy. The above equation appears to be a reasonable description of the temperature dependence of the linewidth of the A exciton in pristine $MoS_2$ and $WS_2$ as shown in Fig. SI 15. However, Fig. SI 3 clearly shows that the FWHM versus temperature data in the alloyed $Mo_xW_{1-x}S_2$ cannot be fitted to the above equation. Presumably, in the alloyed semiconductor, the use of an average phonon energy is not an adequate representation of the electron-phonon interactions. Perhaps both acoustic as well as optical phonon interactions have to be taken into account with the former dominating at low to intermediate temperature range and the latter at high temperatures[3].

**Temperature dependence of intensity for alloyed monolayer $Mo_xW_{1-x}S_2$**

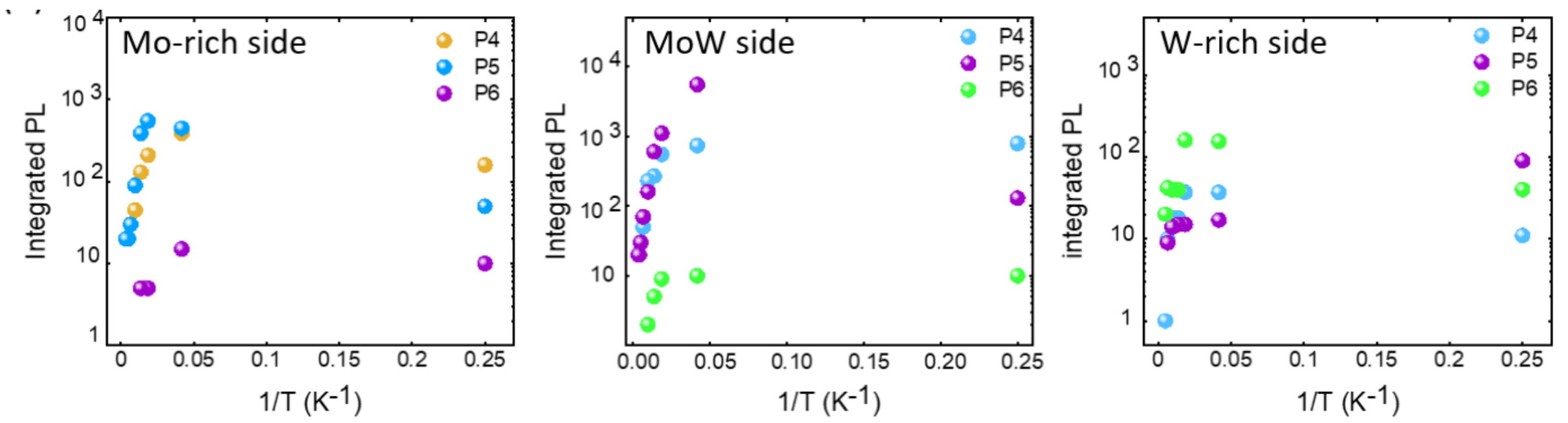

**Figure SI 4**. Temperature dependence of intensity for far from the band edge peaks (peaks 4-6) of alloyed monolayer $Mo_xW_{1-x}S_2$.

**Temperature dependence of PL from different regions of alloyed monolayer $Mo_xW_{1-x}S_2$**

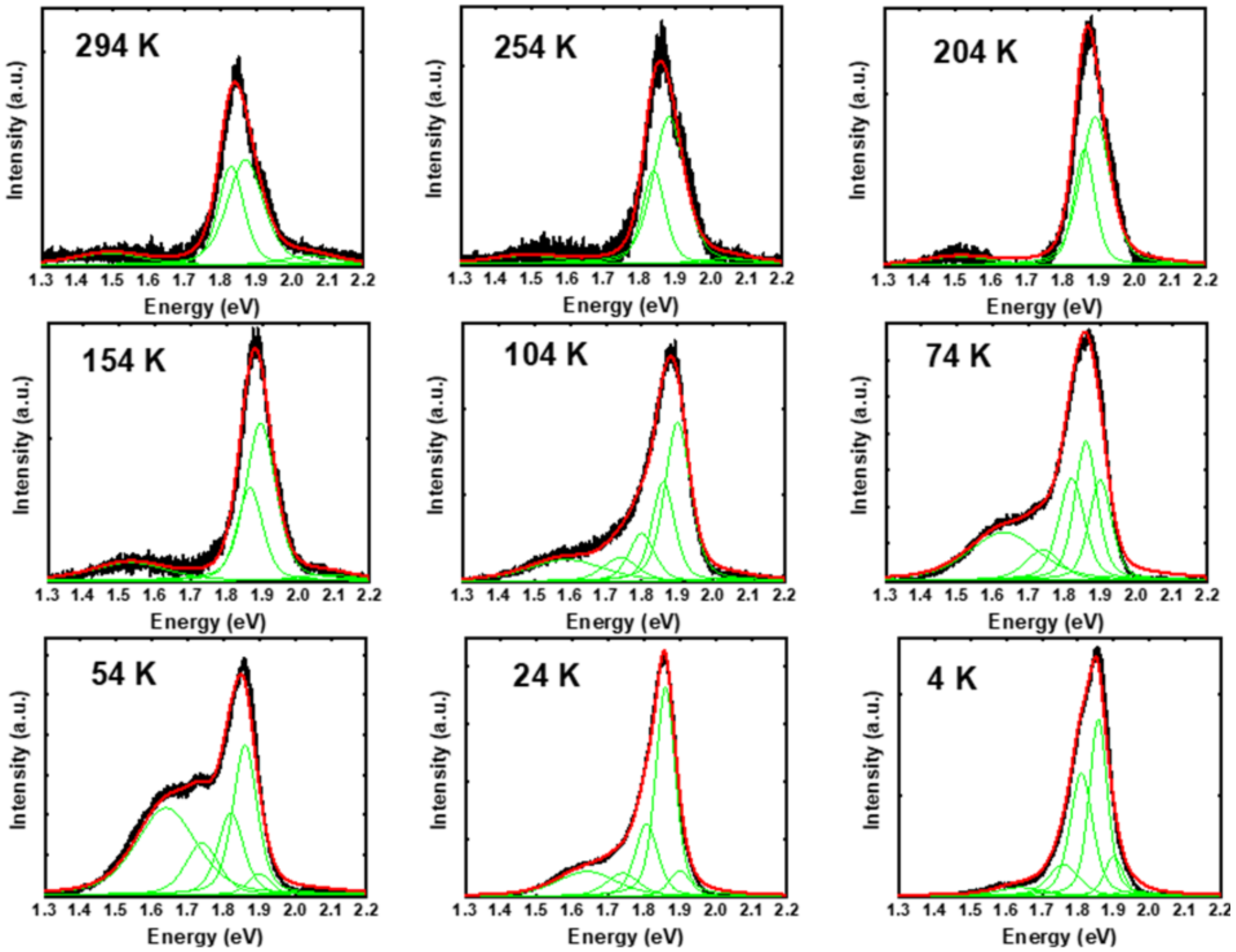

**Figure SI 5**. PL spectra from Mo-rich side of alloyed $Mo_xW_{1-x}S_2$ at different temperatures ranging from 4 K to 300 K. The spectra are deconvoluted with six pseudo-Voigt peaks below 104 K and four peaks above 104 K.

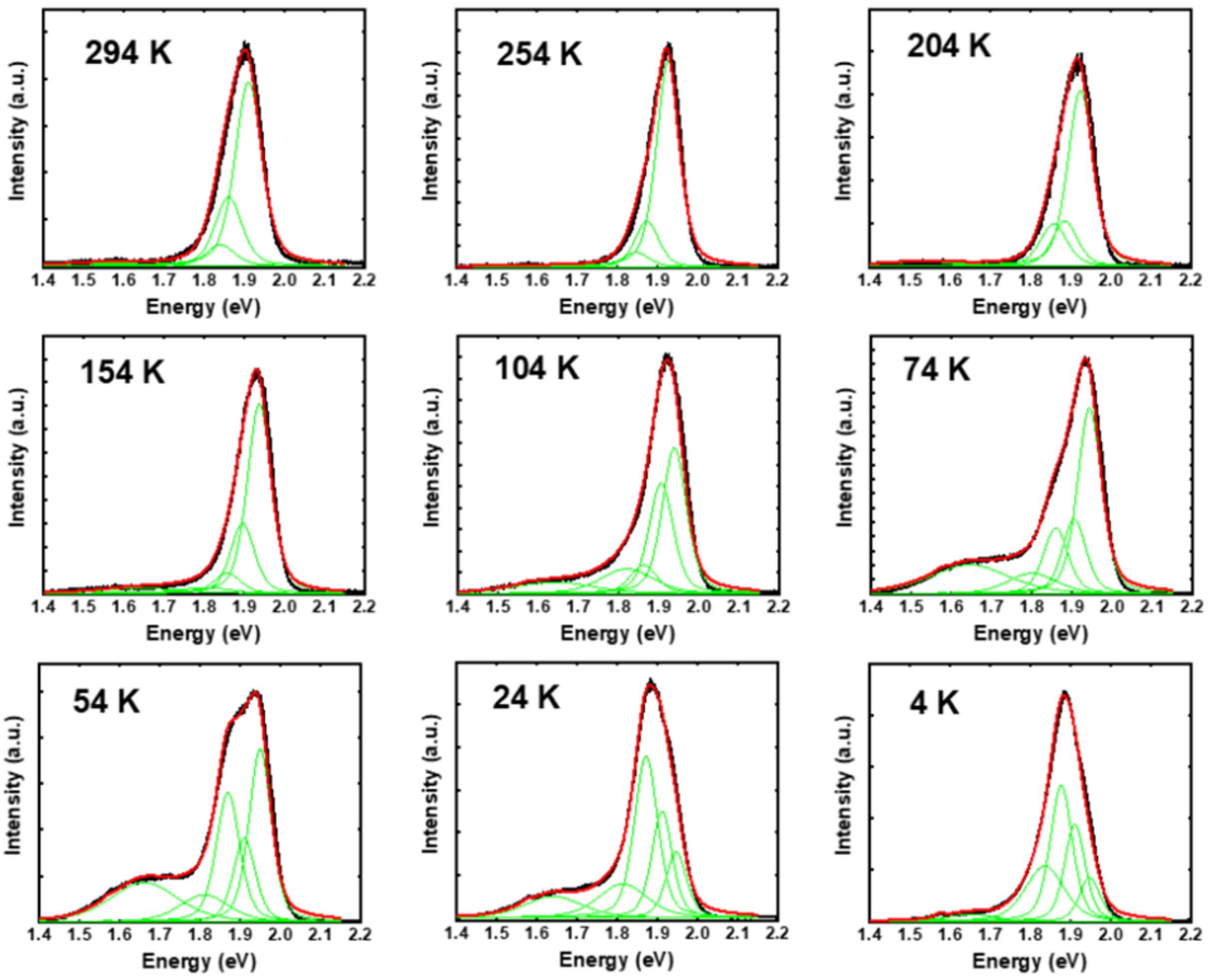

**Figure SI 6**. PL spectra from intermediate region (MoW-side) of alloyed $Mo_xW_{1-x}S_2$ at different temperatures ranging from 4 K to 300 K. The spectra are deconvoluted with six pseudo-Voigt peaks below 104 K and four peaks above 104 K.

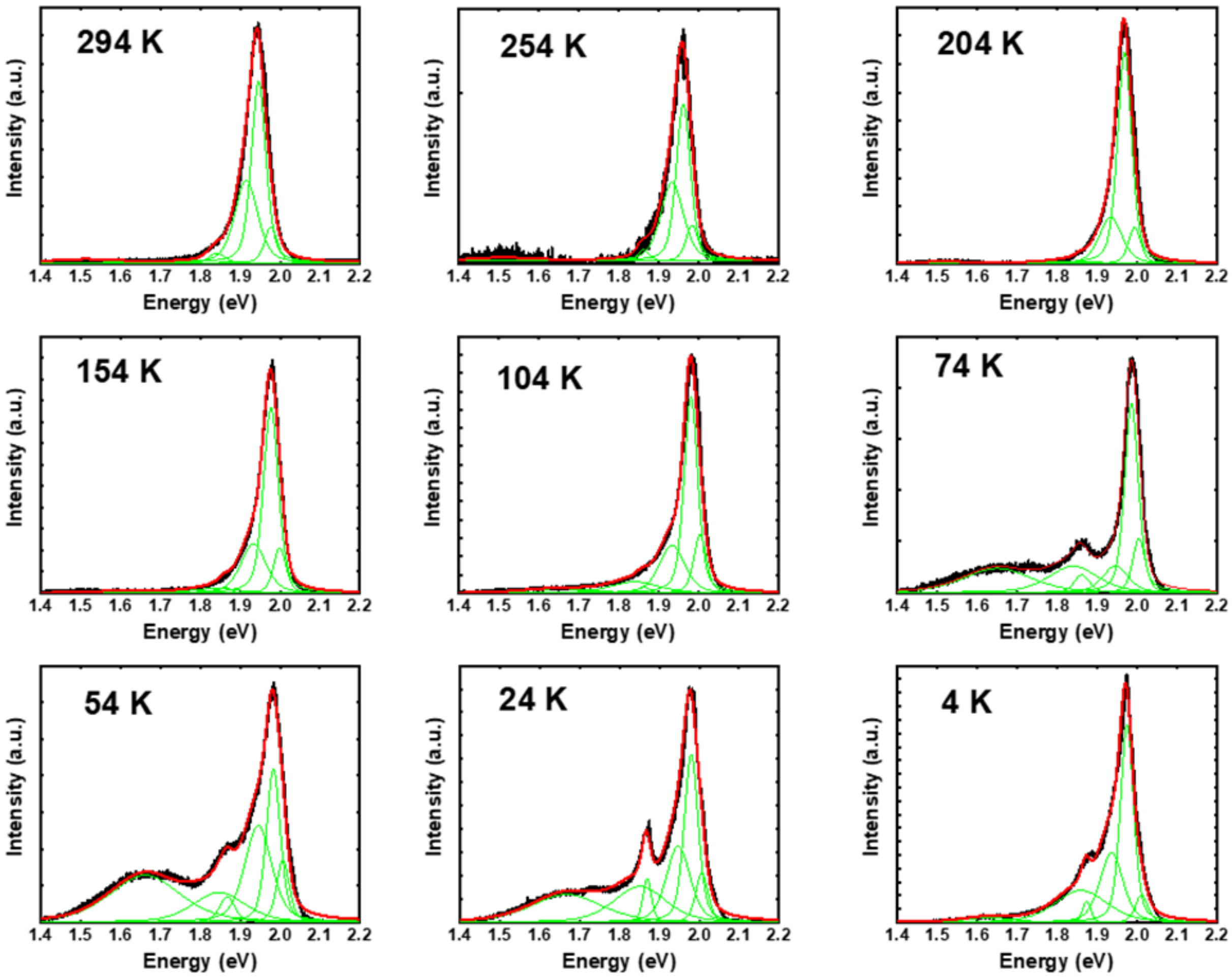

**Figure SI 7**. PL spectra from W-rich side of alloyed $Mo_xW_{1-x}S_2$ at different temperatures ranging from 4 K to 300 K. The spectra are deconvoluted with six pseudo-Voigt peaks below 104 K and five peaks above 104 K.

**Comparison of PL spectra at low T and room T of different regions of alloyed monolayer $Mo_xW_{1-x}S_2$**

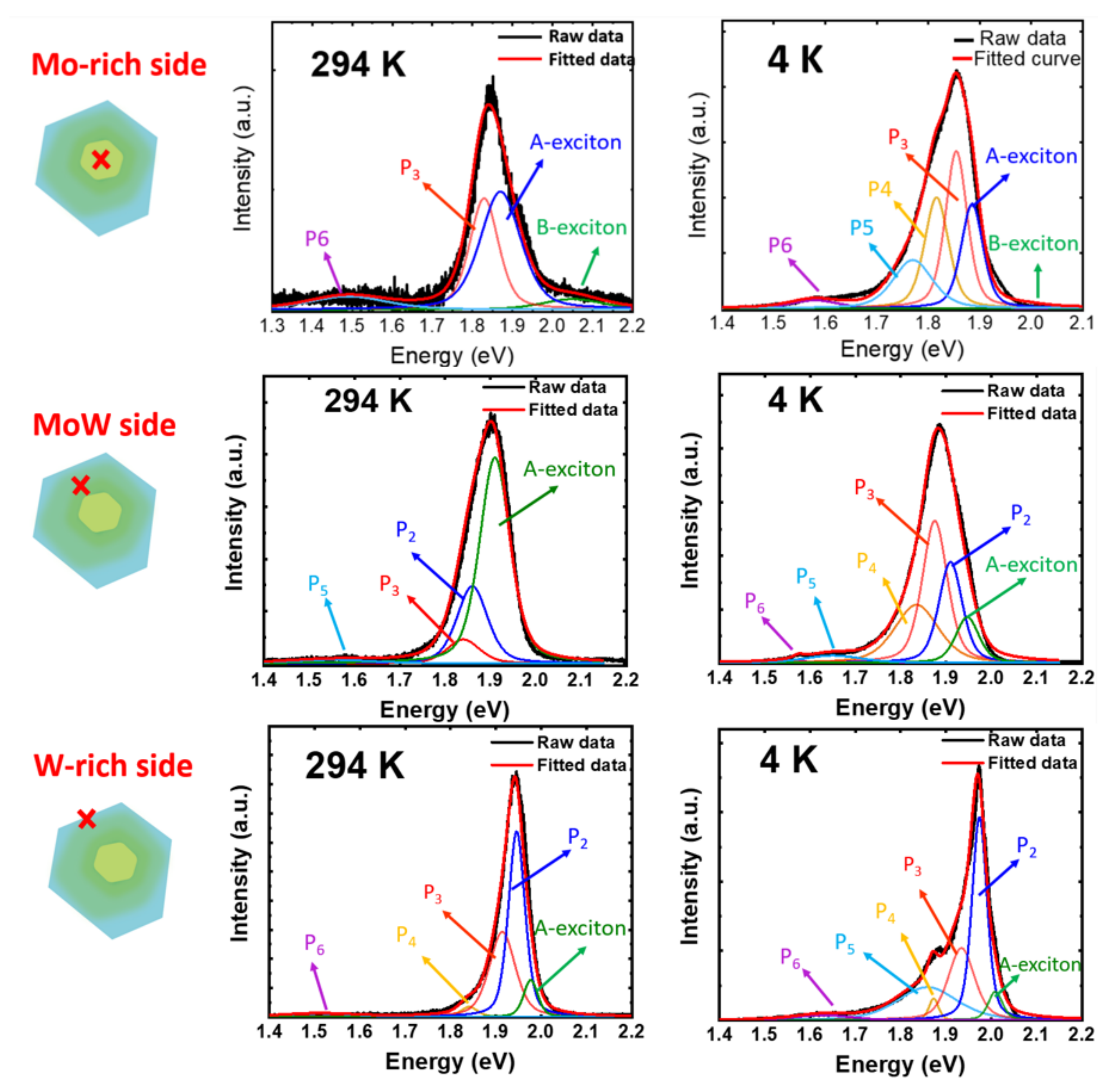

**Figure SI 8**. Comparison of PL spectra at low T and room T of three different regions of alloyed $Mo_xW_{1-x}S_2$.

## Power dependence for Pristine $MoS_2$ and pristine $WS_2$ at low T

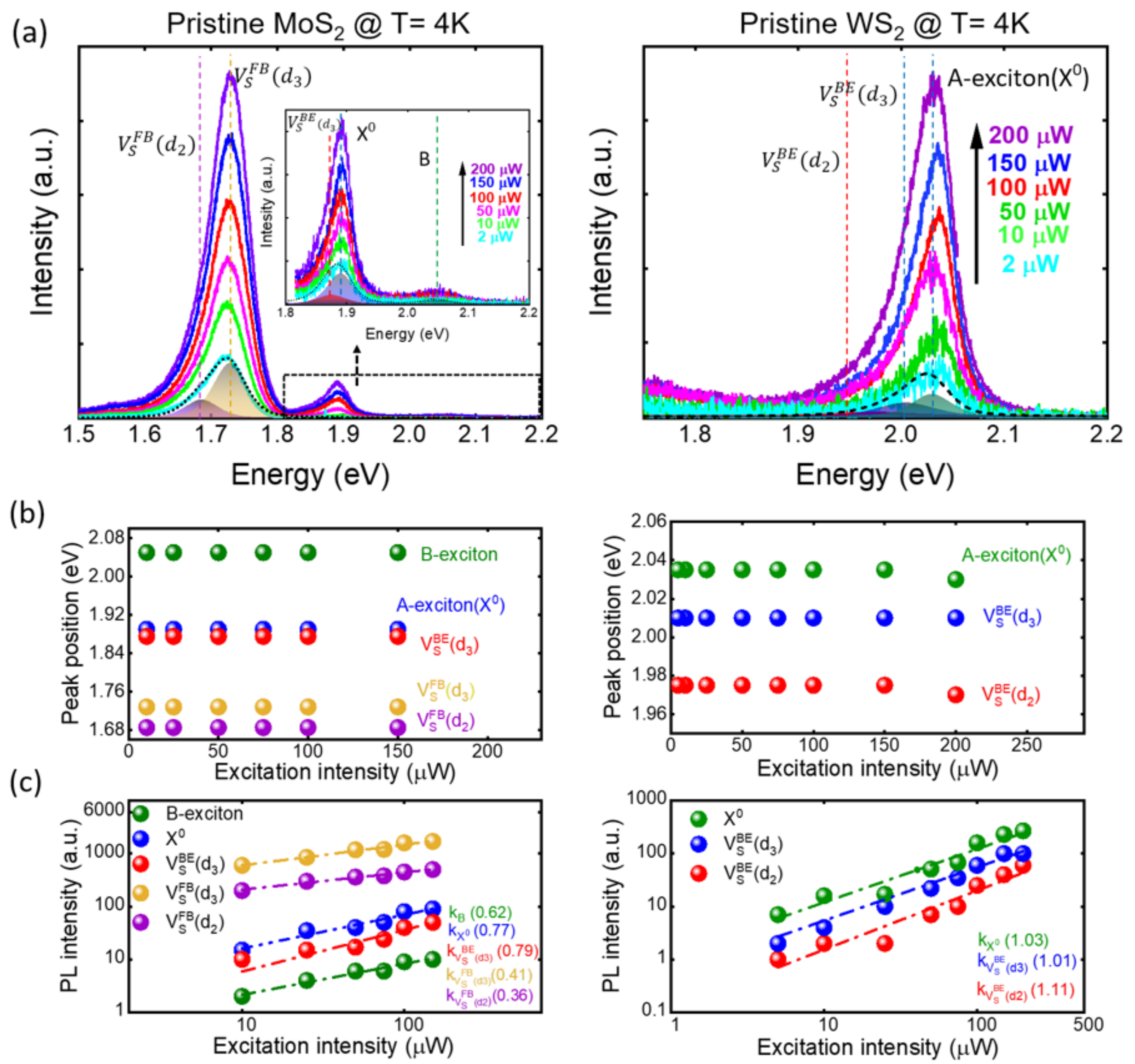

**Figure SI 9**. Power dependence for pristine $MoS_2$ and pristine $WS_2$ monolayers at T = 4 K. (a) Shows the dependence spectra at different excitation powers. (b) Power dependence of peak position for fitted data of PL spectra. (c) Intensity of emission peaks at different excitation powers for fitted data of PL spectra.

To understand the effect of the alloy on exciton emission and benchmark the alloy spectra, we performed power-dependent PL spectroscopy on pristine $MoS_2$ and $WS_2$ that were grown using identical CVD approach to that of the alloyed ones used above. The experiments were performed at low temperature (T = 4 K) and similar laser excitation parameters (wavelength and power) as in the case of for $Mo_xW_{1-x}S_2$ were used. Figure SI 9 shows the PL spectra and analysis for both pristine $WS_2$ and $MoS_2$ acquired using excitation laser power from 2 µW to 200 µW. For pristine $WS_2$, the PL spectra were deconvoluted using three pseudo-Voigt spectral shapes. We identify the three peaks with neutral A-exciton (2.04 eV), $V_S^{BE}(d_3)$ at 2.01 eV, and $V_S^{BE}(d_2)$ at 1.95 eV as summarized in Table 1 in the main article. A similar analysis using five pseudo-Voigt spectral shapes for $MoS_2$ identifies the intrinsic excitonic peaks, B exciton (2.05 eV), and A exciton (1.895 eV), and a bound exciton associated with $V_S$ level $d_3$, $V_S^{BE}(d_3)$ (1.875 eV) (Table 1 of the main article). It is interesting to note that the free-to-bound transitions associated with $V_S$ ($V_S^{FB}(d_3)$ *and* $V_S^{FB}(d_2)$) dominate the radiative process relative to the intrinsic excitons at low temperature. Similar dominant low energy transitions that have been observed previously were attributed to excitons bound to defects or impurities.[4,5] . However, in this study, based on the observed sublinear excitation dependence of the peaks, they are assigned to free-to-bound transitions from the doublet $V_S$ states.

**Power dependence of FHWM for pristine $MoS_2$ and pristine $WS_2$ at T = 4 K**

We have plotted the FWHM of different fitted data for both pristine $MoS_2$ and $WS_2$ (Fig. SI 10). in $MoS_2$. The FWHM of A-exciton is considerably broader than what is reported for exfoliated ML $MoS_2$ [6]. The FWHM of A-exciton(~40 meV) for pristine $WS_2$ at low temperature is considerably broader than what has been reported for exfoliated ML $WS_2$ (@ 7K, 18 meV) [6] and also broader than CVD grown $WS_2$ previously reported value of 23 meV at 77K [7]. In both $MoS_2$ and $WS_2$ the broadening can be attributed to inhomogeneous broadening caused perhaps by various factors such as doping, defects, strain, substrate effects.

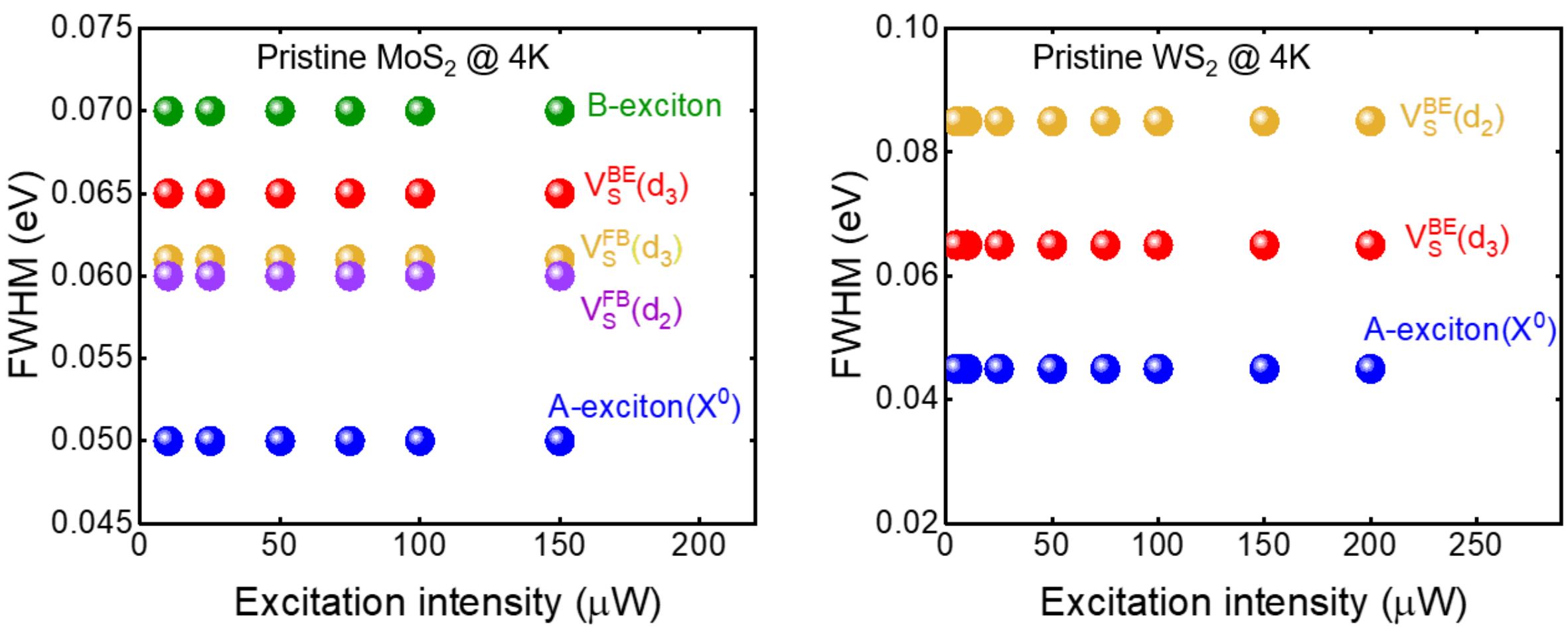

**Figure SI 10**. Power dependence of FWHM for different peaks of pristine $MoS_2$ and $WS_2$ showing negligible effect.

**Power dependence for Pristine $MoS_2$ and pristine $WS_2$ at room T**

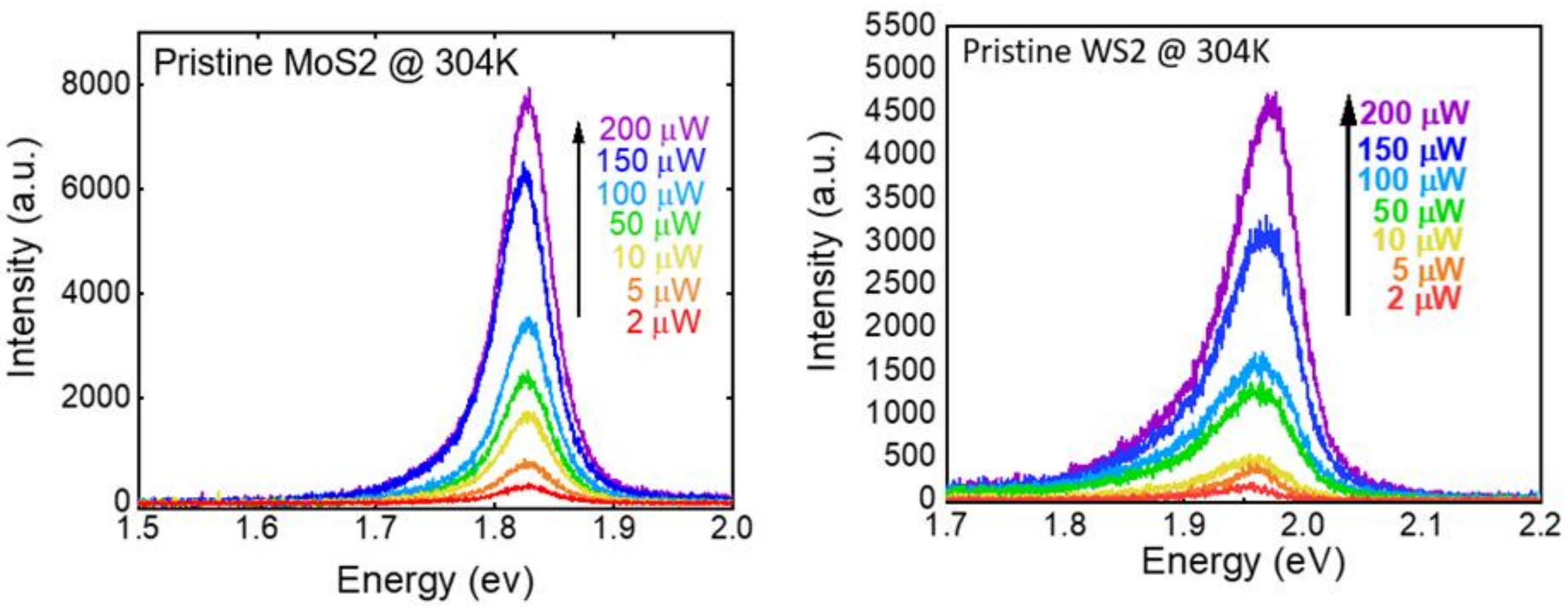

**Figure SI 11.** Power dependence of pristine $WS_2$ and $MoS_2$ at T = 304 K.

**Temperature dependence of PL for pristine $MoS_2$ and $WS_2$**

We also performed temperature-dependent PL spectroscopy of the pristine $MoS_2$ and $WS_2$ samples. The experiments were performed in the temperature range of 4 K to 304 K. Figures SI 12 and SI 13show the PL spectra and analysis of temperature dependence for pristine and $MoS_2$ and $WS_2$, respectively. At room temperature (T = 304 K), fitting the spectral shape to pseudo-Voigt function yields the peak positions of the B-exciton at 2.01 eV and A-exciton ($X^0$) at 1.83 eV. Two other peaks at 1.79 eV and 1.71 eV are attributed to bound excitons associated with the doublet state of $V_S$ (see DFT calculations)[8]. As the temperature is decreased, at 104K there is the evolution of a broad band at 1.7 eV which decomposes to two peaks at 4K which are attributed to free-to-bound transitions from the same doublet state of $V_S$ wherein a photoexcited electron is captured at the defect level and recombines radiatively with a hole in the valence band. Due to the reduced thermal ionization of the captured electron at low temperatures, the free-to-bound transitions, labelled as $V_S^{FB}(d_3)$ and $V_S^{FB}(d_2)$, are prominent at 4K. In fact, they are the dominant transitions at 4K relative to the excitonic transitions. The intensity of the A-exciton remains almost constant as a function of temperature as shown in Fig. SI 15. For pristine $WS_2$, at room temperatures, the PL spectra can be fitted using three peaks at 1.963 eV (A-exciton), and two bound exciton transitions peaks $V_S^{BE}(d_3)$ and $V_S^{BE}(d_2)$ with energy levels at 1.897 eV , and 1.85 eV (Fig. SI 13) which are in a good agreement with reported values[9-11] .

As temperature decreases, the PL peaks show a blueshift. Figure SI 14 (b) shows the PL peak positions at different temperatures for both pristine $MoS_2$ and $WS_2$. The temperature dependence of the PL peaks can be fitted in accordance with Eq. (2) shown in the main article. In Table SI 3 are shown the fitting parameters for the A-exciton, $V_S^{BE}(d_3)$ and $V_S^{BE}(d_2)$ bound excitons for pristine $MoS_2$ and $WS_2$, and the values of the average phonon energies are comparable with that of alloy $Mo_xW_{1-x}S_2$.

**Temperature dependence of PL for pristine $MoS_2$**

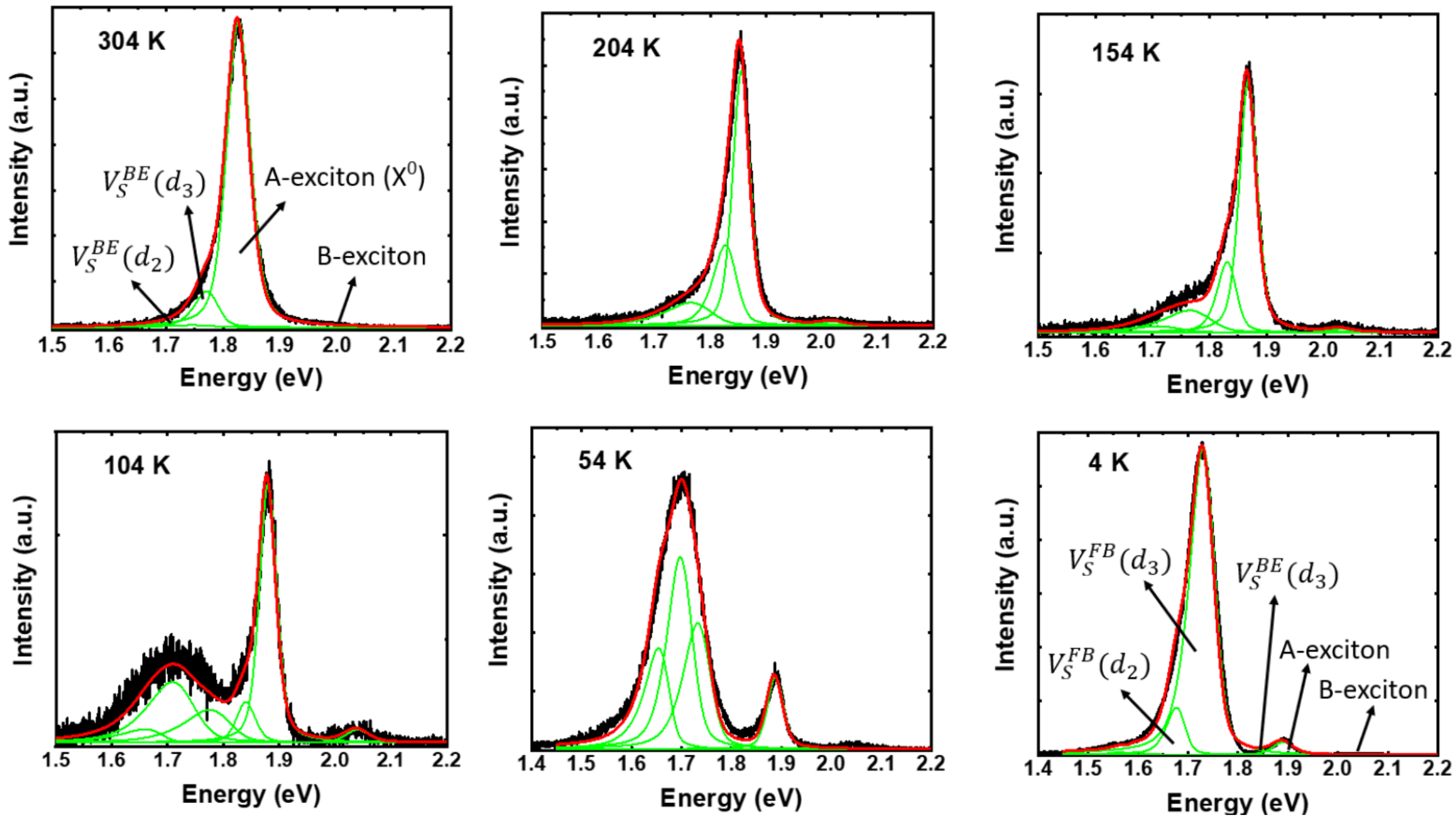

**Figure SI 12**. PL spectra and spectral fitting of pristine $MoS_2$ for temperature ranging from 4K-304K.

## Temperature dependence of PL for pristine $WS_2$

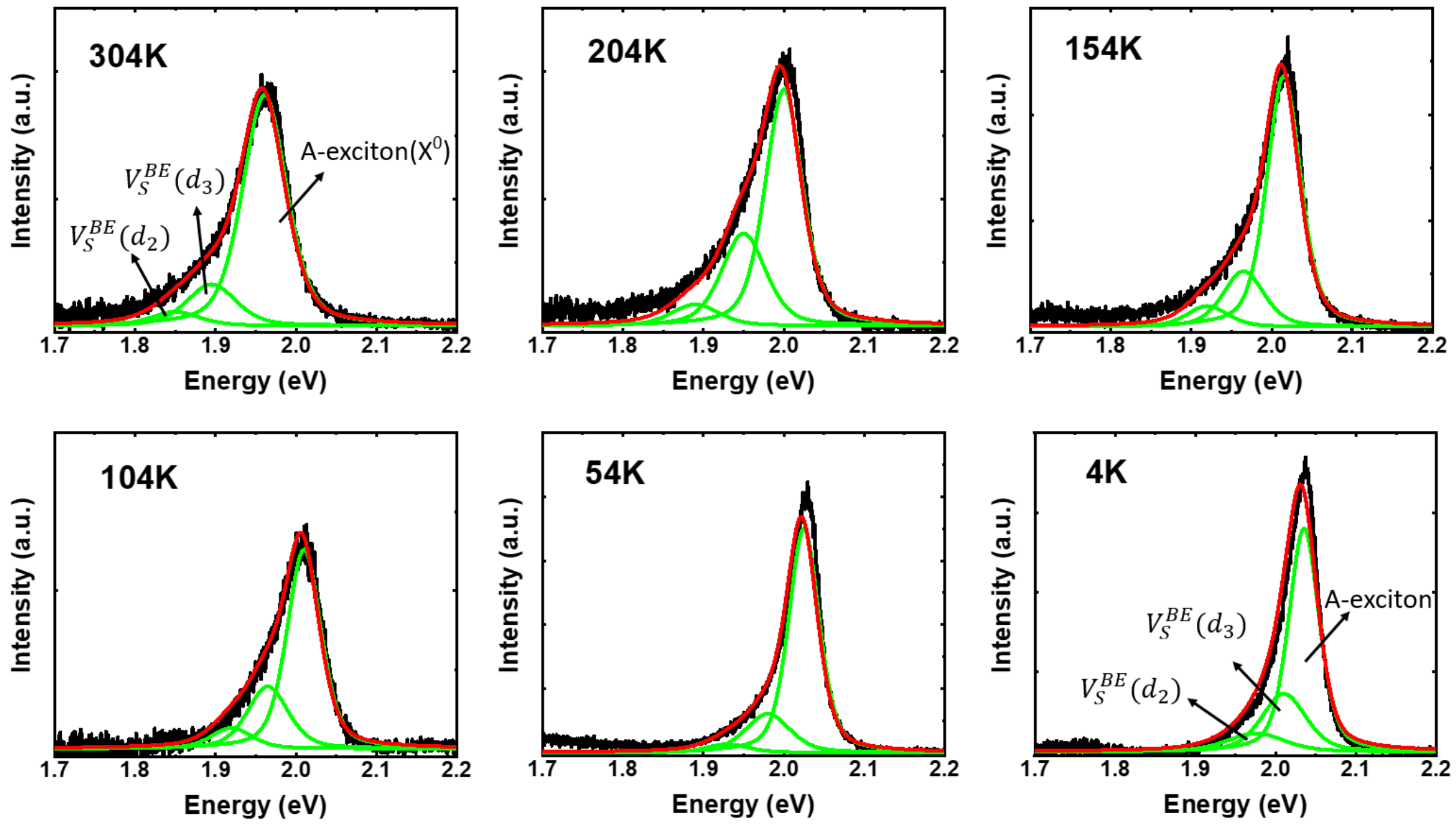

**Figure SI 13**. PL spectra and spectral fitting of pristine $WS_2$ for temperature ranging from 4K-304K.

## Comparison of temperature dependence of PL spectra for pristine $MoS_2$ and pristine $WS_2$

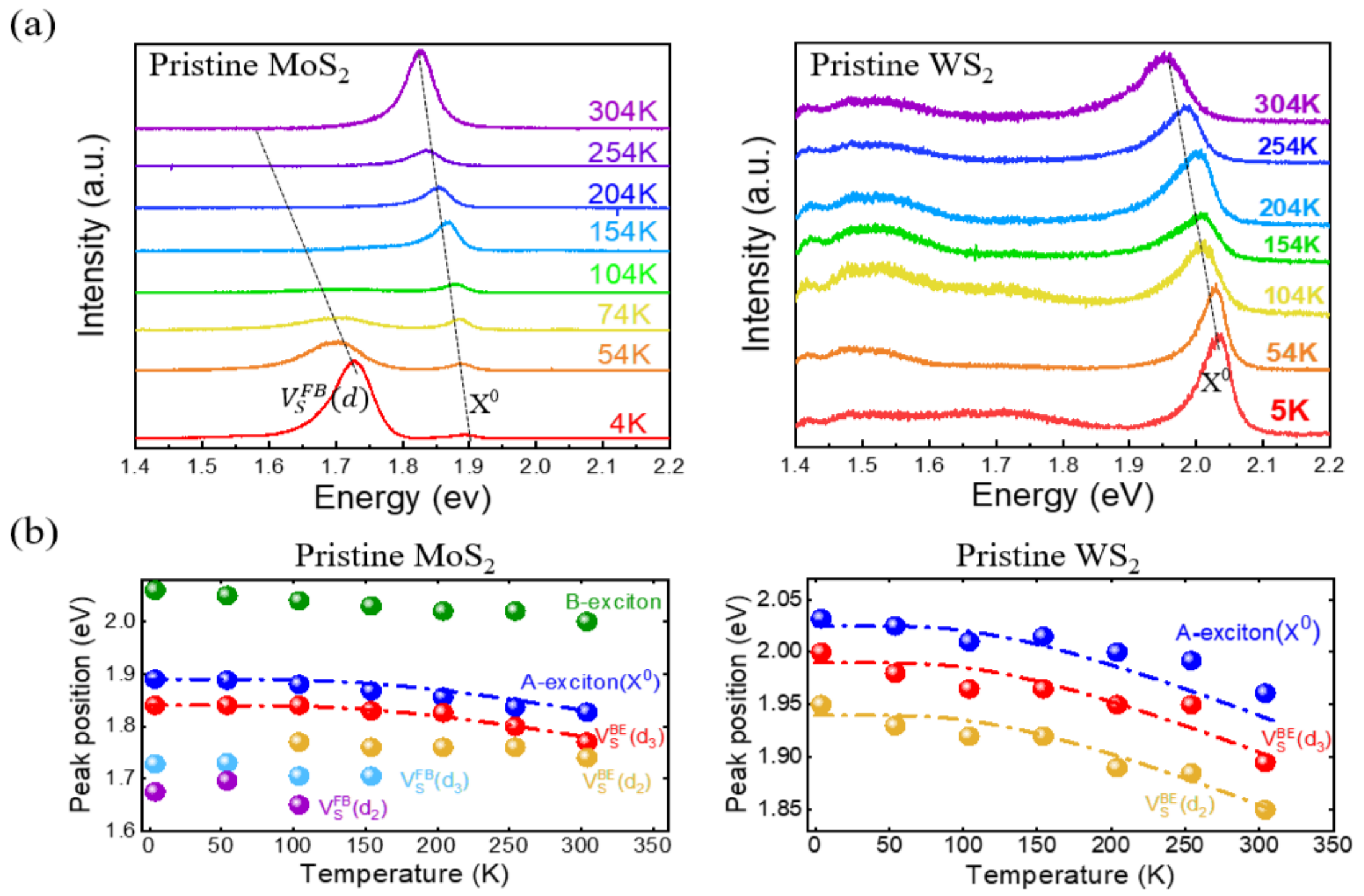

**Figure SI 14**. (a) The temperature dependence of PL spectra from T = 300 K to 4 K for both pristine $MoS_2$ and $WS_2$. (b) The temperature dependence of peak positions for both pristine $MoS_2$ and $WS_2$. Dash-dot lines are fitting curves using Eq. (2) in the main article.

**Temperature dependence of FWHM for pristine $MoS_2$, $WS_2$**

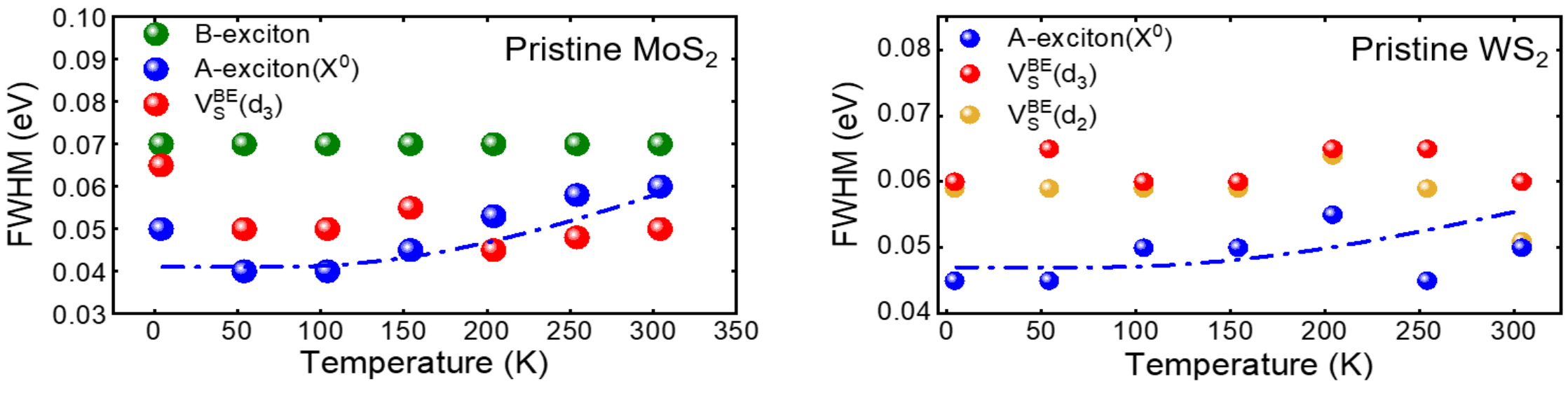

**Figure SI 15**. Temperature dependence of FWHM for different peaks of pristine $MoS_2$ and $WS_2$ with dash-dot curved lines are fitted data using Eq. (1).

**Temperature dependence of intensity for pristine $MoS_2$, $WS_2$**

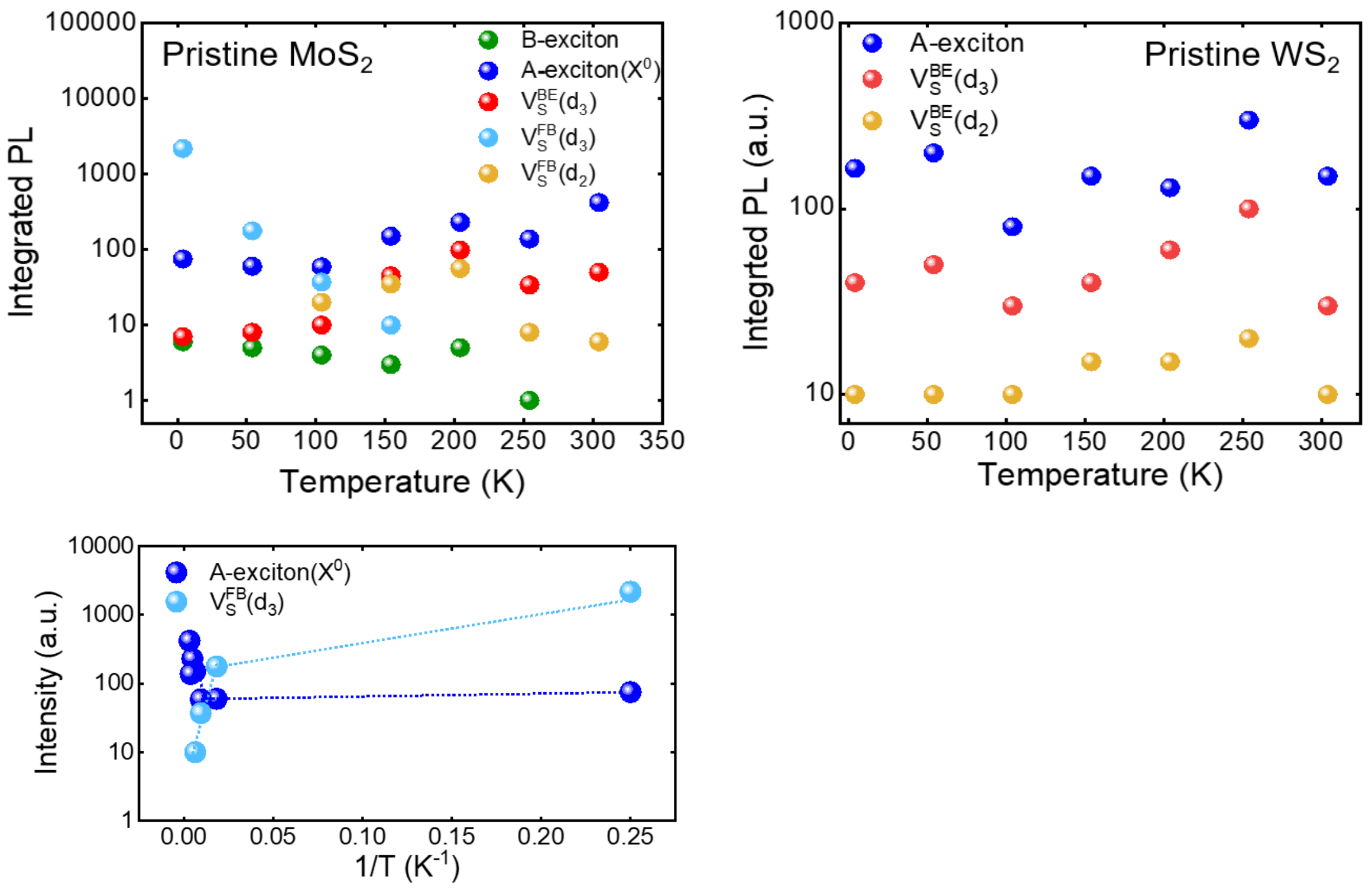

**Figure SI 16**. Temperature dependence of Intensity for different peaks of pristine $MoS_2$ and $WS_2$.

**Compilation of peak energies of A-exciton and B-exciton of monolayer $WS_2$ and $MoS_2$.**

Optical transitions from monolayer $WS_2$ and $MoS_2$have been rigorously studied by several groups at low temperature as well as room temperature. The observed peak position of A-exciton varies over a wide range from ~ 1.90 eV to 1.96 eV in $MoS_2$ (Tables SI 2) and from ~ 2.01 eV to 2.10 eV in $WS_2$ (Table SI 1). On the low energy side of the A-exciton an optical transition in the range 1.88 eV to 1.92 eV in $MoS_2$ and 1.96 eV to 2.05 eV in $WS_2$ is typically attributed to $A^-$ trion. Since the intensity of the optical transition in these above energy ranges show linear dependence on excitation intensity as opposed to a $3/2$ dependence expected for a three particle trion transition, the involvement of a trion in the radiative process is not unambiguous. It is to be noted that optical transitions on the low energy side of the A-exciton have also been assigned to defect/impurity related bound excitons (See Tables SI 1 and SI 2). In this work we have attributed the low energy transitions to $V_S$ defects.

**Table SI 1**. The summary of PL transition energies associated with $A^-$ trion, A-exciton, and defects in monolayer $WS_2$.

| Article | $A^-$ trion (eV) | A-exciton (eV) | B-exciton (eV) | Defect (eV) |
|---|---|---|---|---|
| Krustok et.al [15] @ 10K_CVD | 1.97 | 2.02 | - | 1.94 |
| Wan et. al [16] @ 4K_CVD | 2.02 | 2.08 | - | 1.98 |
| Kaupmees [17] et. al @15K_CVD | 1.96 | 2.01 | - | - |
| Plechinger et.al [11] @ 4K_exfoliated | 2.04 | 2.09 | - | - |
| Jadczak et.al [6] @ 7K_exfoliated | 2.05 | 2.10 | - | - |
| This work @ 4K (pristine $WS_2$) | - | 2.04 | - | - |
| This work @ 4K (alloyed monolayer $Mo_xW_{1-x}S_2$, W-rich side) | - | 2.01 | - | 1.94-1.871.86-1.63 |

**Table SI 2**. The summary of PL transition energies associated with $A^-$ trion, A-exciton, B-exciton and defects in monolayer $MoS_2$.

| Article | $A^-$ trion (eV) | A-exciton (eV) | B-exciton (eV) | Defect (eV) |
|---|---|---|---|---|
| Mak et.al [12] @ 10K | 1.90 | 1.92 | - | - |
| Jadczak et.al [6] @ 7K_exfoliated | 1.92 | 1.96 | - | - |
| Cadiz et.al [18] @ 10K | 1.92 | 1.96 | - | - |
| Pei et.al [19] @ 10K | 1.88 | 1.92 | - | - |
| Sharma et.al [13] @ 4K_CVD | 1.88 | 1.92 | 2.08 | 1.80-1.87 |
| Verhagen et.al [4] @ 10K_CVD | 1.87 | 1.89 | 1.96 | 1.73-1.78 |
| Panday et.al [14] @ 4K_exfoliated | 1.92 | 1.96 | 2.08 | 1.82 |
| Christopher et.al [20] @ 83K | 1.92 | 1.95 | 2.09 | -- |
| This work @ 4K (pristine $MoS_2$) | - | 1.895 | 2.05 | 1.73-1.68 |
| This work @ 4K (alloyed monolayer $Mo_xW_{1-x}S_2$, Mo-rich side) | - | 1.89 | 2.01 | 1.82-1.77-1.58 |

**Table SI 3**. Fitted Values of the Exciton−Phonon Coupling Strength, S, the Average Phonon Energy, ⟨ħω⟩, and $E_{g(0)}$ of neutral A exciton, $V_S^{BE}(d_3)$ and $V_S^{BE}(d_2)$ for different regions of alloyed monolayer $Mo_xW_{1-x}S_2$ compared to pristine $MoS_2$ and $WS_2$.

| | Pristine $MoS_2$ | Pristine $WS_2$ | Mo-rich side of alloyed $Mo_xW_{1-x}S_2$ | MoW side of alloyed $Mo_xW_{1-x}S_2$ | W-rich side of alloyed $Mo_xW_{1-x}S_2$ |
|---|---|---|---|---|---|
| $E_{0A}$ (eV) | 1.89 | 2.04 | 1.90 | 1.96 | 2.01 |
| $\boldsymbol{V_S^{BE}(d_3)}$ (eV) | 1.84 | 1.99 | | 1.91 | 1.98 |
| $\boldsymbol{V_S^{BE}(d_2)}$ (eV) | | 1.94 | 1.862 | 1.875 | 1.95 |
| <ħω> (meV) | 50 | 35 | 50 | 35 | 44 |
| S | 3.5 | 3.5 | 1.91 | 2.2 | 2.2 |

**Raman and PL Spectra at room temperature of alloyed $Mo_xW_{1-x}S_2$ monolayer**

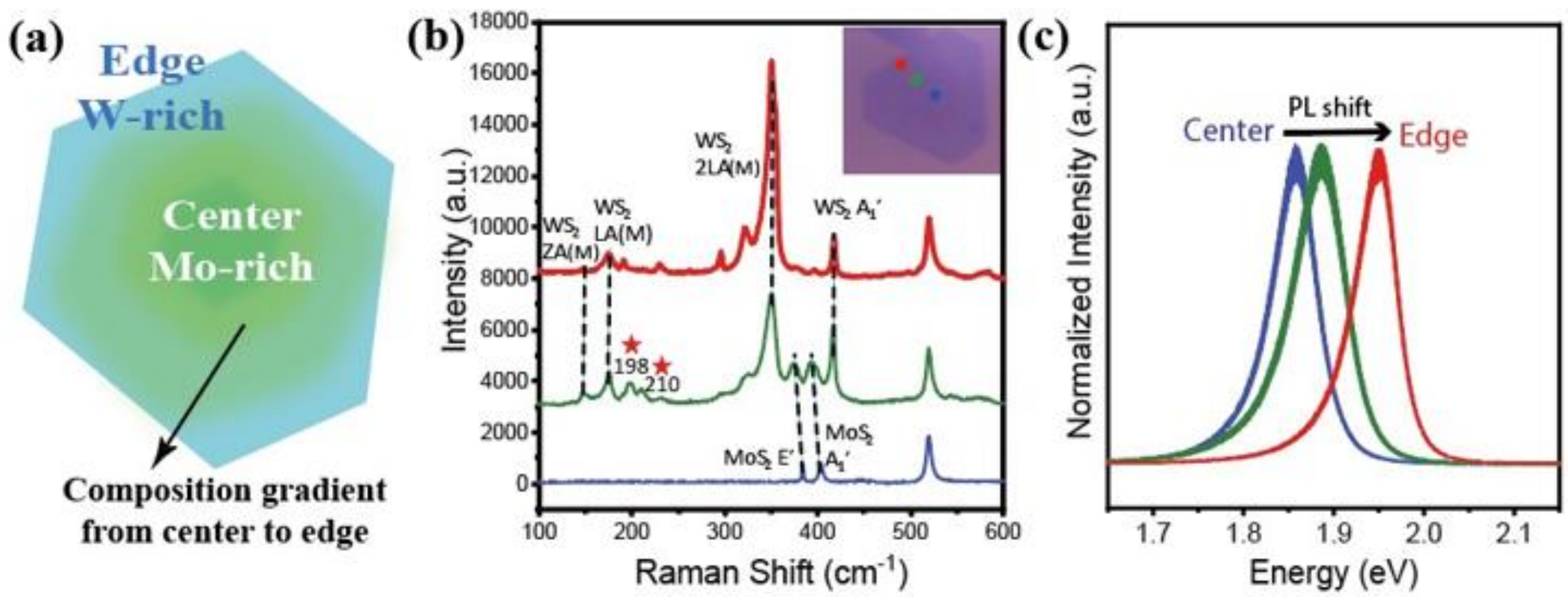

**Figure SI 17**. Far-Field Optical Characterization of alloyed monolayer $Mo_xW_{1-x}S_2$. (a) Schematic of the structure of alloyed monolayer $Mo_xW_{1-x}S_2$. (b) Raman and (c) PL spectra of various regions in alloyed monolayer $Mo_xW_{1-x}S_2$. The excitation laser is at 532 nm.

**DFT calculations**

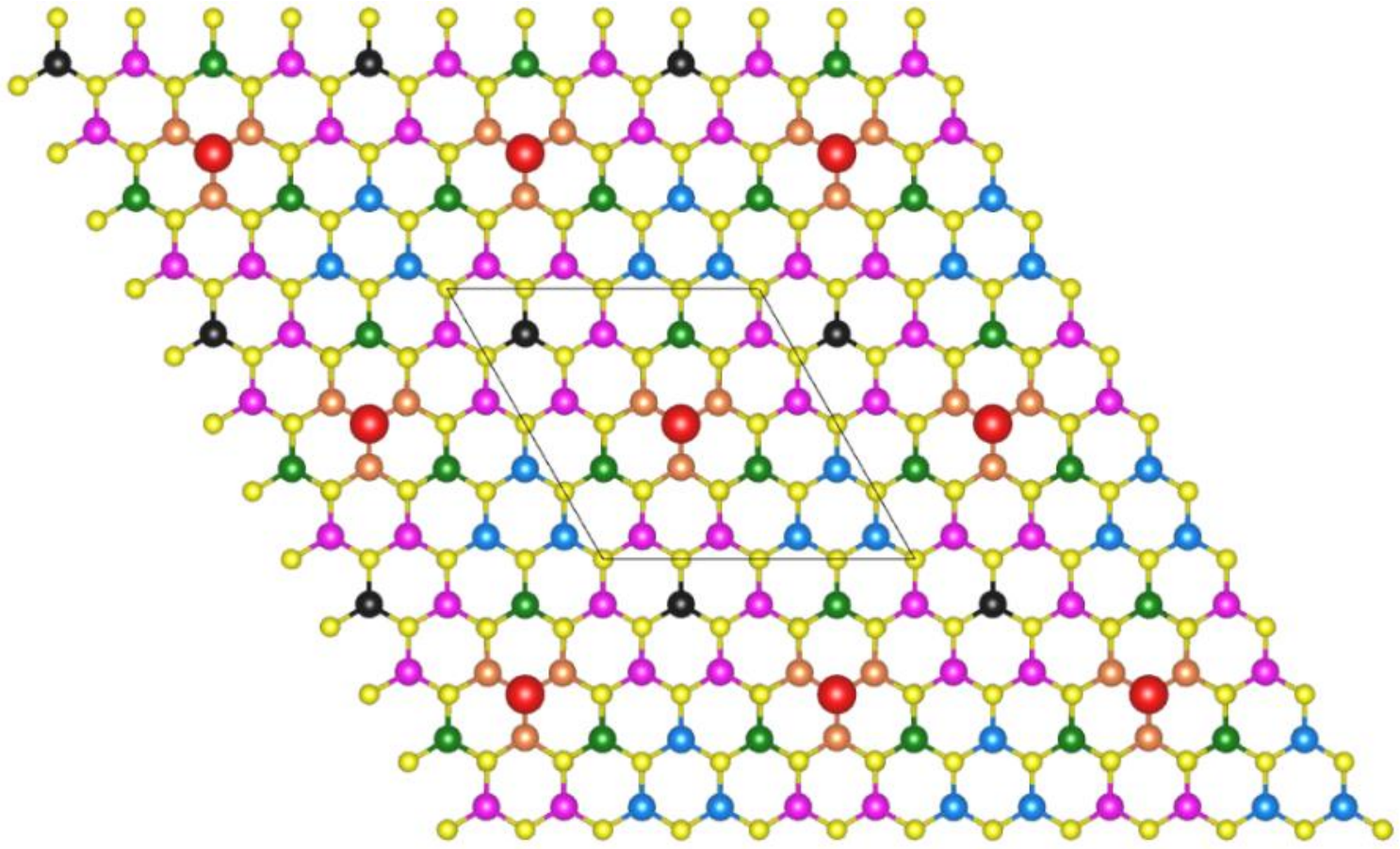

**Figure SI 18.** The 4 x 4 x 1 $Mo_{16x}W_{16(1-x)}S_{31}$ supercell (bounded by black lines) illustrating all possible $C_{3v}$ symmetric structures. The large red atoms are situated at the position of the S vacancies to emphasize their placement and the other colors (orange, green, pink, teal, and black) represent the position that a transition metal (Mo or W) can be situated to generate the 32 possible structures (5 distinct colors with 2 possible atoms each yields $2^5 = 32$ structures).

For all transitions, the corrected energy ($E_{g(A,B)}^{corr}$) for the alloys is determined from the equation:

$$E_{g(A,B)}^{corr} = E_g[\frac{E_{EXP(A,B)}^{MoS_2}}{E_{PBE(A,B)}^{MoS_2}} x + \frac{E_{EXP(A,B)}^{WS_2}}{E_{PBE(A,B)}^{WS_2}} (1 - x)] \qquad (2)$$

where $x$ is the Mo concentration and $E_g$ is the PBE calculated A (B) exciton transition energy, $E_{PBE(A,B)}^{MoS_2}$= 1.615 eV (1.759 eV) and $E_{PBE(A,B)}^{WS_2}$= 1.607 eV (2.001 eV) are the PBE calculated transition energies for $Mo_{16}S_{32}$ and $W_{16}S_{32}$, respectively, for the A (B) exciton energies. Similarly, $E_{EXP(A,B)}^{MoS_2}$= 1.88 eV (2.02 eV) and $E_{EXP(A,B)}^{WS_2}$= 2.02 eV (2.40 eV) are the experimental A (B) exciton energies for pristine $Mo_{16}S_{32}$ and $W_{16}S_{32}$, respectively [21,22]. For defect-related transitions, the same correction applied to the A [B] exciton transitions, is applied to the spin-down (red-red, $v$-$d_2$, $v$-$d_3$) [spin-up, (blue-blue, $v_i$-$d_1$, $v_i$-$d_4$)] transitions. The valence band splitting can be represented by the linear relationship described by Vegard's law [23] which is expressed in terms of the Mo concentration ($x$) as

$$E_v(x) = xE_{v,Mo_{16}S_{31,32}} + (1 - x)E_{v,W_{16}S_{31,32}} \qquad (3)$$

where $E_{v,Mo_{16}S_{31,32}}$ and $E_{v,W_{16}S_{31,32}}$ represent the valence ($v$) band splitting for non-defective ($Mo_{16}S_{32}$ and $W_{16}S_{32}$) and $V_S$ defective ($Mo_{16}S_{31}$ and $W_{16}S_{31}$) transition metal dichalcogenides. In the cases of the A and B exciton transition energies and conduction band splitting, the relationship becomes quadratic due to the introduction of a bowing parameter ($b$), which is described by the relationship:

$$E_{g,c}(x) = xE_{g,c,Mo_{16}S_{31}} + (1-x)E_{g,c,W_{16}S_{31}} - bx(1-x) \quad (4)$$

where $E_{g,Mo_{16}S_{31,32}}$ and $E_{g,W_{16}S_{31,32}}$ are the transition energies for $Mo_{16}S_{31,32}$ and $W_{16}S_{31,32}$, respectively, and similarly $E_{c,Mo_{16}S_{31,32}}$ and $E_{c,W_{16}S_{31,32}}$ are the conduction band splitting for $Mo_{16}S_{31,32}$ and $W_{16}S_{31,32}$, respectively. This analysis reveals a strong linear correlation for the valence band splitting as a function of the Mo concentration, but there is bowing observed for the A and B excitons and the conduction band splitting (Fig. 5 and Fig. SI 20). As the Mo concentration increases in the pristine [$Mo_{16x}W_{16(1-x)}S_{32}$] alloys, the A and B exciton decreases quadratically (see Fig. 5). The defect-mediated transitions are shown in an example with the band structure of an $Mo_7W_9S_{31}$ with the $v$-$d_2$, $v$-$d_3$, $v_i$-$d_1$, and $v_i$-$d_4$ transitions which are referred to as $R_1$, $R_2$, $B_1$, and $B_2$, respectively (Fig. SI 19).

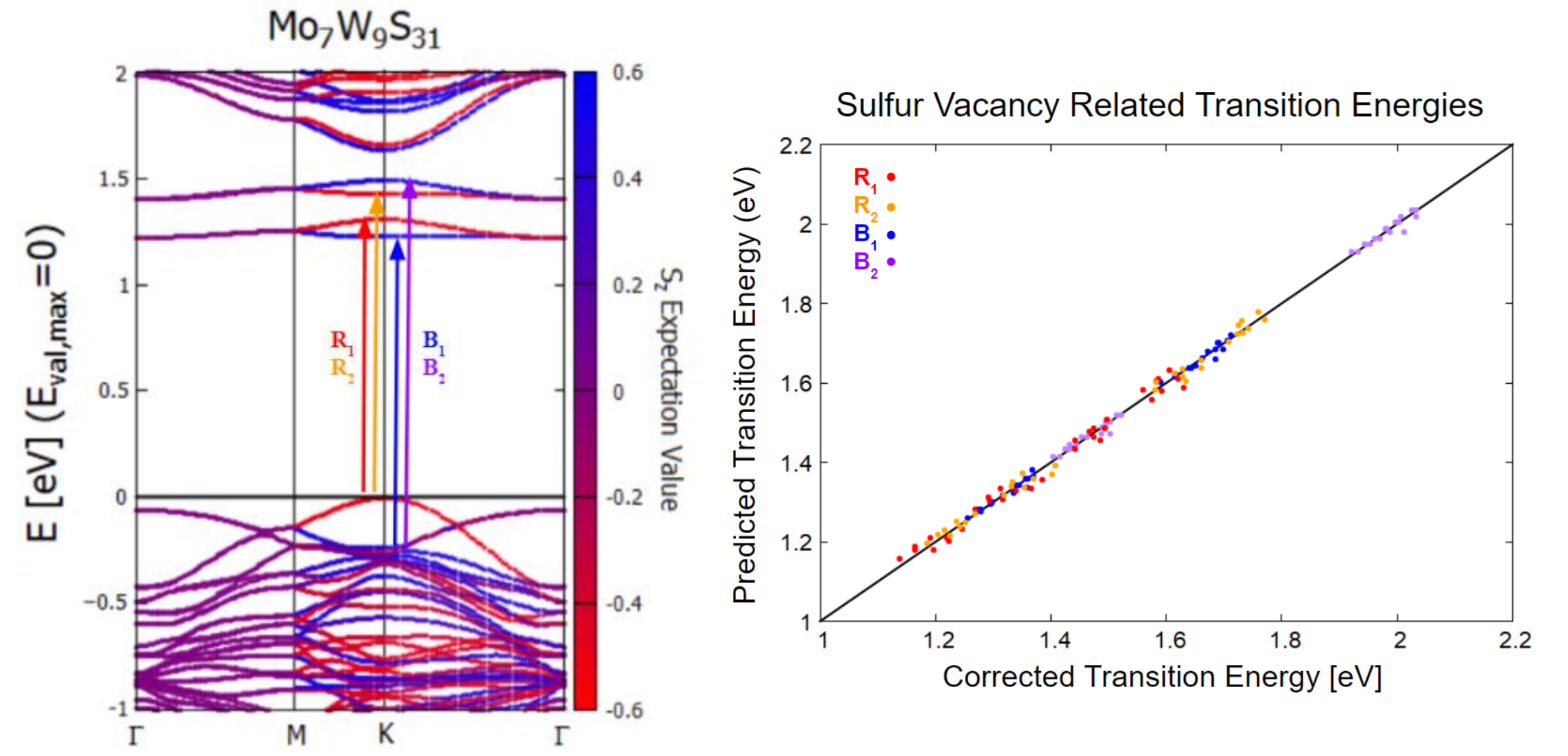

**Figure SI 19**. The $Mo_7W_9S_{31}$ band structure calculated using the PBE functional is used as an example to illustrate the position of the defect-mediated energy levels in the alloy [left]. In this case, only the $R_1$ and $R_2$ and $B_1$ and $B_2$ defect-mediated transitions are shown. [right] The predicted defect-related transition energies are plotted against the shifted energy from the energies calculated using the PBE functional with the identity line (essentially f(x) = x) shown as a black line.

In order to create a model to predict the defect-mediated transition energies, a parameter ($t_i$) is introduced where $t_i$ = -1 or 1 based on which atom (Mo or W) occupies the corresponding-colored position (from Fig. SI 18) where $t_i$ = -1 for Mo atoms and $t_i$ = 1 for W atoms. The best fit equation for this model is applied to the five sets of colored atoms [black (B), pink (P), teal (T), green (G), and orange (O)] to create a linear combination, which is represented by

$$E_{trans} = b + a_B t_B + a_P t_P + a_T t_T + a_G t_G + a_O t_O \quad (5)$$

where $b$, $E_{trans}$, and $a_i$ represent a constant, particular transition ($B_1$, $R_1$, $R_2$, and $B_2$), and weighting coefficient corresponding to a specific color ($i$) from Fig. SI 19, respectively. Based on the 32 possible structures investigated, the best-fitting linear combinations generated for the $B_1$, $R_1$, $R_2$, and $B_2$ transitions are

$$E_{B1} = 1.491 - 0.008t_B - 0.031t_P + 0.011t_T + 0.010t_G + 0.171t_O \quad (6)$$

$$E_{R1} = 1.395 - 0.011t_B - 0.063t_P - 0.015t_T - 0.011t_G + 0.138t_O \quad (7)$$

$$E_{R2} = 1.487 - 0.011t_B - 0.061t_P - 0.010t_T - 0.016t_G + 0.193t_O \quad (8)$$

$$E_{B2} = 1.725 - 0.029t_P + 0.009t_T + 0.015t_G + 0.258t_O \quad (9)$$

In general, the placement of atoms near $V_S$ (or orange position) has the strongest effect with the placement of W atoms tending to increase the defect-mediated transition energies. For all cases regarding the defect-mediated transitions, the presence of W atoms at the pink and black positions tends to decrease the transition energy. For the spin-down (red) transitions, the presence of W atoms at the green and teal positions tends to decrease the transition energies, while for the spin-up (blue) transitions, the presence of W atoms at the green and teal positions tends to increase the transition energies. Additionally, the difference between the $d_2$ and $d_3$ energy levels becomes significantly larger when $V_S$ is adjacent to W atoms, but when the vacancy is adjacent to Mo atoms, then the difference becomes significantly smaller (Fig. SI 21). Overall, this model predicts the defect-mediated transition energies very well with strong correlation ($R^2$ = 0.996) with predictions residing very close to the identity line.

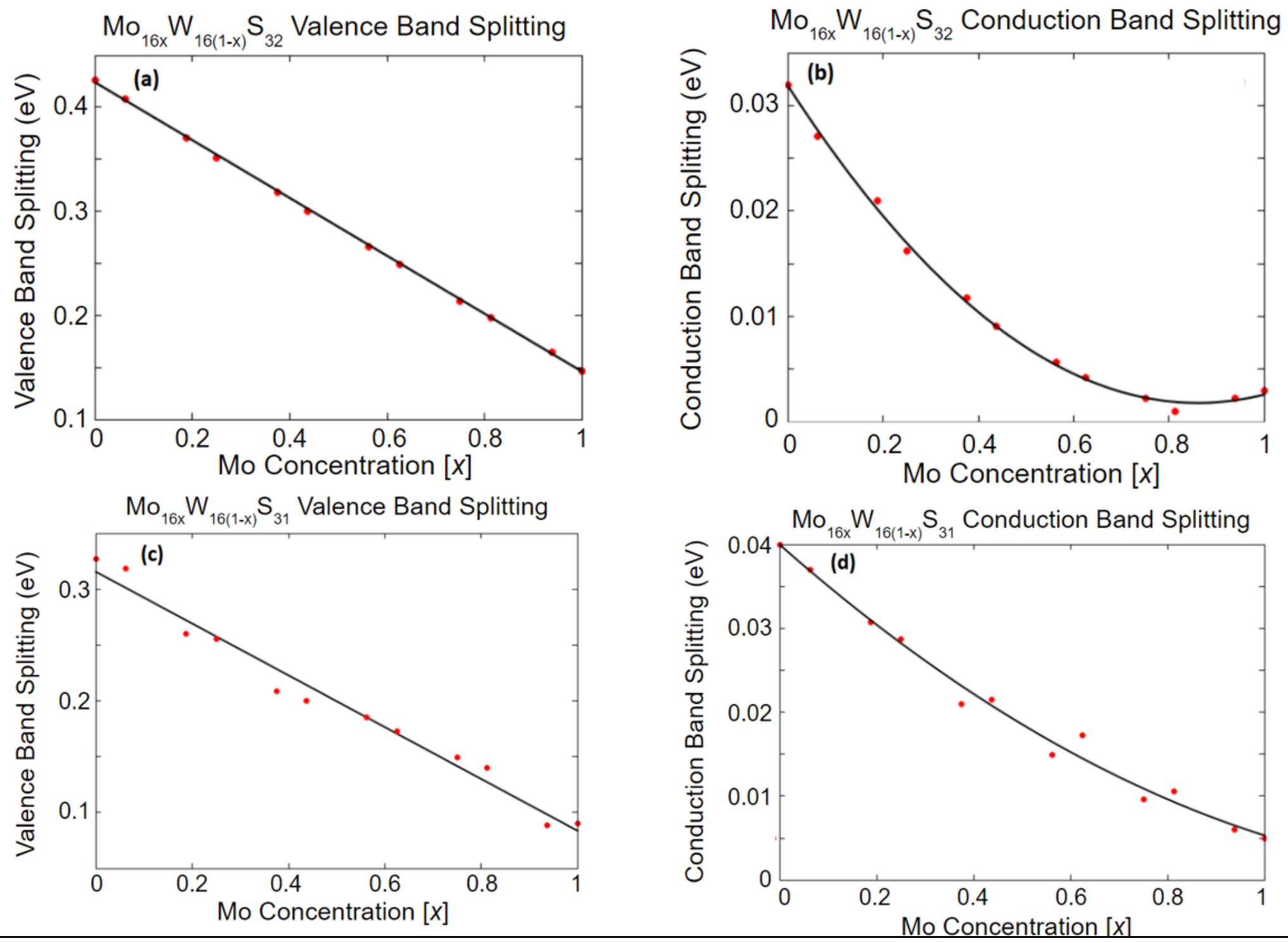

**Figure SI 20**. The values for the (a) valence and (b) conduction band splitting due to SOC in $Mo_{16x}W_{16(1-x)}S_{32}$ and the (c) valence and (d) conduction band splitting in $Mo_{16x}W_{16(1-x)}S_{31}$ as a function of Mo concentration [*x*].

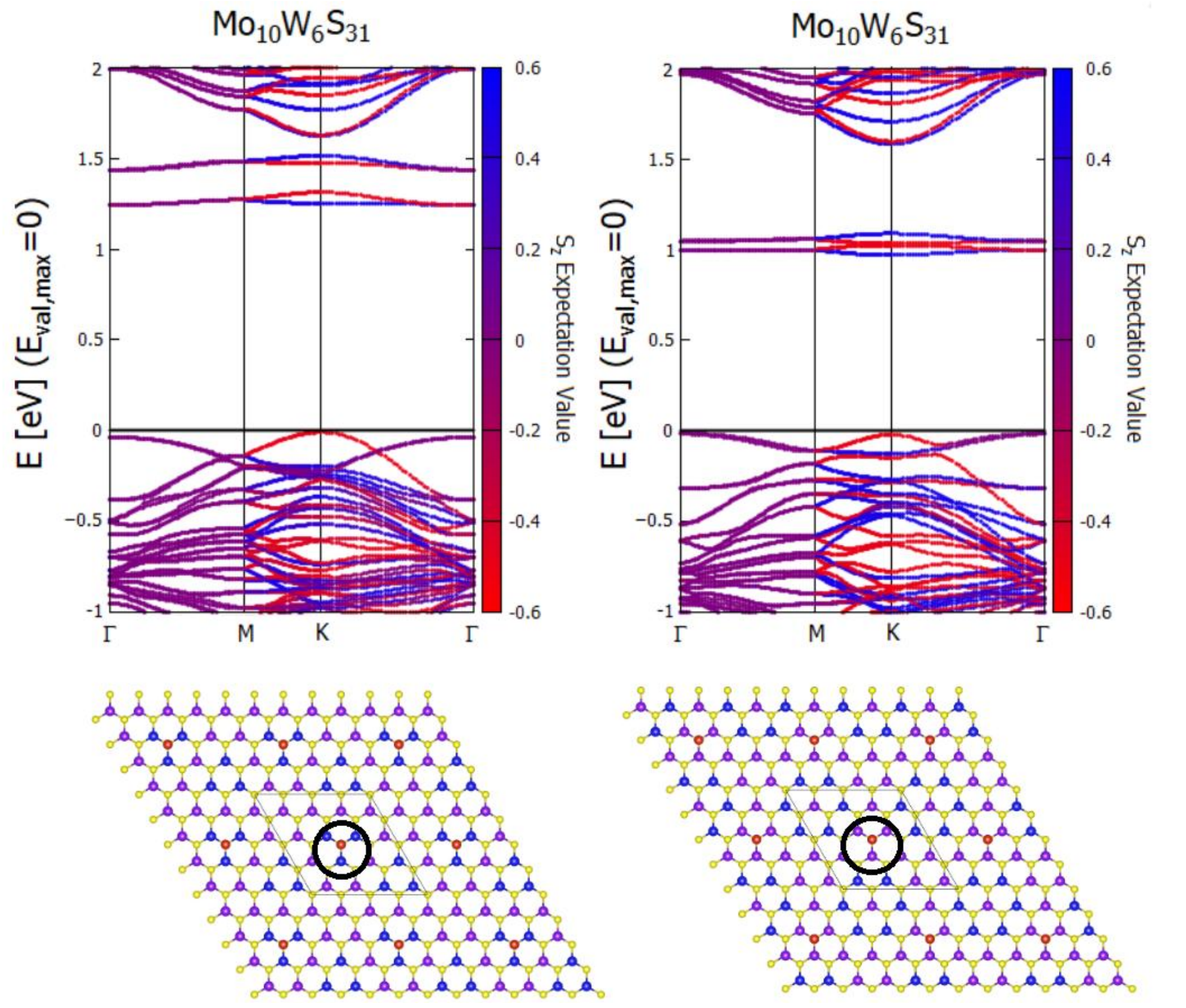

**Figure SI 21.** [Top] Two $Mo_{10}W_6S_{31}$ band structures calculated using the PBE functional where the W atoms surround the $V_S$ defect (left) and the Mo atoms surround the $V_S$ defect (right). The left configuration shows shallow defect levels, while the right configuration shows deep defect levels. [Bottom] The corresponding $Mo_{10}W_6S_{31}$ configurations where the red atoms emphasize the $V_S$ defect and the purple and blue atoms represent Mo and W atoms, respectively. The black circle shows the $V_S$ defect is surrounded by W atoms [left] and Mo atoms [right].

**Table SI 4**. The averages for the corrected transition energies [$R_1$, $R_2$, $B_1$, and $B_2$] for both when the $V_S$ defect is surrounded by W atoms (W) and by Mo atoms (Mo) for each the W-rich ($x \leq 0.25$), MoW (intermediate, $0.25 < x < 0.75$), and Mo-rich ($x \geq 0.75$) regions. The experimentally observed peaks far from the band edge (P5, P6) that best match the transitions $R_1$ and $R_2$ are shown, respectively, as $R_1$(exp) and $R_2$(exp). Since the transitions involving the spin-orbit split valance band such as the B-excitons are generally weaker than the A-excitons (as in $MoS_2$) or not even observed (as in $WS_2$), no assignment is made to the transitions $B_1$ and $B_2$ involving the spin-orbit split valence band (see Fig. SI 20).

| Region | $R_1$ | $R_2$ | $B_1$ | $B_2$ | $R_1$ (exp) | $R_2$ (exp) |
|---|---|---|---|---|---|---|
| **W-rich (W)** | 1.44-1.49 | 1.58-1.66 | 1.62-1.69 | 1.94-2.01 | - | P6 |
| **MoW (W)** | 1.48-1.63 | 1.63-1.77 | 1.59-1.71 | 1.92-2.03 | P6 | P5 |
| **Mo-rich (W)** | 1.59-1.61 | 1.73-1.76 | 1.66-1.67 | 1.98 | P6 | P5 |
| **W-rich (Mo)** | 1.14-1.16 | 1.19-1.20 | 1.30-1.32 | 1.45-1.46 | - | - |
| **MoW (Mo)** | 1.17-1.29 | 1.22-1.33 | 1.25-1.37 | 1.40-1.52 | - | - |
| **Mo-rich (Mo)** | 1.29-1.39 | 1.33-1.41 | 1.33-1.36 | 1.49-1.50 | - | - |

## Supplementary References